\documentstyle[12pt,epsfig]{article}
\newlength{\dinwidth}
\newlength{\dinmargin}
\begin{document}
%
\newcommand {\Mchi}        {M_\chi}
\newcommand {\Mchip}       {M_{\chi^\prime}}
\newcommand {\Mcha}        {M_{\chi^\pm}}
\newcommand {\Msnu}        {M_{\tilde{\nu}}}
\newcommand {\snu}         {\tilde{\nu}}
\newcommand {\cha}         {\chi^\pm}
\newcommand {\chip}        {\chi^\prime}
\newcommand {\chipp}       {\chi^{\prime\prime}}
\newcommand {\chippp}      {\chi^{\prime\prime\prime}}
\newcommand {\mhalf}       {m_{1/2}}
\newcommand {\tb}          {\tan\beta}
\newcommand {\ssqtw}       {\sin^2\theta_{\mathrm{W}}}
\newcommand {\ALEPH}       {{\sc ALEPH}}
\newcommand {\lep}         {{\sc LEP}}
\newcommand {\lepa}        {{\sc LEP~1}}
\newcommand {\lepb}        {{\sc LEP~1.5}}
\newcommand {\goto}        {\rightarrow}
\newcommand {\GeV}         {{\mathrm{GeV}}}
\newcommand {\GeVc}        {{\mathrm{GeV}}/c}
\newcommand {\GeVcsq}      {{\mathrm{GeV}}/c^2}
\newcommand {\etal} {{\it et al }}
%
\newcommand{\appgt}{\, \raisebox{-0.5ex}{$\stackrel{\scriptstyle >}%
 {\scriptstyle\sim}$}\,}
\newcommand{\applt}{\, \raisebox{-0.5ex}{$\stackrel{\scriptstyle <}%
 {\scriptstyle\sim}$}\,}
\newcommand {\onehalf} {\raisebox{0.1ex}{${\frac{1}{2}}$}}
\thispagestyle{empty}
~~\\
\begin{center}
\centerline{EUROPEAN LABORATORY FOR PARTICLE PHYSICS (CERN)}
%
\vglue 0.4cm
\begin{flushright}
CERN-PPE/96-083\\
July 2, 1996
\end{flushright}
\vskip 2cm
{\Huge Mass Limit for the Lightest Neutralino}\\
\vspace{1cm}
{\Large The ALEPH Collaboration}
\vspace{1cm}
\begin{quotation}
\noindent
{\bf Abstract:}
{\small
Indirect limits on the mass of the lightest neutralino are derived from the
results of searches for charginos, neutralinos, and sleptons performed with
data taken by the \ALEPH~Collaboration at centre-of-mass energies near the
Z~peak and at 130~and 136~GeV. Within the context of the Minimal
Supersymmetric Standard Model and when $M_{\snu}\ge200~\GeVcsq$, the bound
$\Mchi>12.8~\GeVcsq$ at the 95\% confidence level applies for any $\tan\beta$.
The impact of lighter sneutrinos is presented in the framework of SUSY grand
unified theories; a massless neutralino is allowed only for a narrow range
of $\tb$, $\mu$, and the scalar mass parameter $m_0$.  Finally, by including
Higgs mass constraints and requiring that radiative electroweak symmetry
breaking occur, more stringent bounds on~$\Mchi$ as a function of $\tb$ are
derived.
}
\end{quotation}
\end{center}
\vspace{3cm}
\centerline{{\em (to be submitted to Zeitschrift f\"ur Physik)}}
\vfill
\pagestyle{empty}
\newpage
\small
%
%
\newlength{\saveparskip}
\newlength{\savetextheight}
\newlength{\savetopmargin}
\newlength{\savetextwidth}
\newlength{\saveoddsidemargin}
\newlength{\savetopsep}
\setlength{\saveparskip}{\parskip}
\setlength{\savetextheight}{\textheight}
\setlength{\savetopmargin}{\topmargin}
\setlength{\savetextwidth}{\textwidth}
\setlength{\saveoddsidemargin}{\oddsidemargin}
\setlength{\savetopsep}{\topsep}
%
%
\setlength{\parskip}{0.0cm}
\setlength{\textheight}{25.0cm}
\setlength{\topmargin}{-1.5cm}
\setlength{\textwidth}{16 cm}
\setlength{\oddsidemargin}{-0.0cm}
\setlength{\topsep}{1mm}
\pretolerance=10000
\centerline{\large\bf The ALEPH Collaboration}
\footnotesize
\vspace{0.5cm}
{\raggedbottom
\begin{sloppypar}
\samepage\noindent
D.~Buskulic,
I.~De Bonis,
D.~Decamp,
P.~Ghez,
C.~Goy,
J.-P.~Lees,
A.~Lucotte,
M.-N.~Minard,
J.-Y.~Nief,
P.~Odier,
B.~Pietrzyk
\nopagebreak
\begin{center}
\parbox{15.5cm}{\sl\samepage
Laboratoire de Physique des Particules (LAPP), IN$^{2}$P$^{3}$-CNRS,
74019 Annecy-le-Vieux Cedex, France}
\end{center}\end{sloppypar}
\vspace{2mm}
\begin{sloppypar}
\noindent
M.P.~Casado,
M.~Chmeissani,
J.M.~Crespo,
M.~Delfino, 
I.~Efthymiopoulos,$^{20}$
E.~Fernandez,
M.~Fernandez-Bosman,
Ll.~Garrido,$^{15}$
A.~Juste,
M.~Martinez,
S.~Orteu,
C.~Padilla,
I.C.~Park,
A.~Pascual,
J.A.~Perlas,
I.~Riu,
F.~Sanchez,
F.~Teubert
\nopagebreak
\begin{center}
\parbox{15.5cm}{\sl\samepage
Institut de Fisica d'Altes Energies, Universitat Autonoma
de Barcelona, 08193 Bellaterra (Barcelona), Spain$^{7}$}
\end{center}\end{sloppypar}
\vspace{2mm}
\begin{sloppypar}
\noindent
A.~Colaleo,
D.~Creanza,
M.~de~Palma,
G.~Gelao,
M.~Girone,
G.~Iaselli,
G.~Maggi,$^{3}$
M.~Maggi,
N.~Marinelli,
S.~Nuzzo,
A.~Ranieri,
G.~Raso,
F.~Ruggieri,
G.~Selvaggi,
L.~Silvestris,
P.~Tempesta,
G.~Zito
\nopagebreak
\begin{center}
\parbox{15.5cm}{\sl\samepage
Dipartimento di Fisica, INFN Sezione di Bari, 70126
Bari, Italy}
\end{center}\end{sloppypar}
\vspace{2mm}
\begin{sloppypar}
\noindent
X.~Huang,
J.~Lin,
Q. Ouyang,
T.~Wang,
Y.~Xie,
R.~Xu,
S.~Xue,
J.~Zhang,
L.~Zhang,
W.~Zhao
\nopagebreak
\begin{center}
\parbox{15.5cm}{\sl\samepage
Institute of High-Energy Physics, Academia Sinica, Beijing, The People's
Republic of China$^{8}$}
\end{center}\end{sloppypar}
\vspace{2mm}
\begin{sloppypar}
\noindent
R.~Alemany,
A.O.~Bazarko,
P.~Bright-Thomas,
M.~Cattaneo,
P.~Comas,
P.~Coyle,
H.~Drevermann,
R.W.~Forty,
M.~Frank,
R.~Hagelberg,
J.~Harvey,
P.~Janot,
B.~Jost,
E.~Kneringer,
J.~Knobloch,
I.~Lehraus,
G.~Lutters,
E.B.~Martin,
P.~Mato,
A.~Minten,
R.~Miquel,
Ll.M.~Mir,$^{2}$
L.~Moneta,
T.~Oest,$^{1}$
A.~Pacheco,
J.-F.~Pusztaszeri,
F.~Ranjard,
P.~Rensing,$^{25}$
L.~Rolandi,
D.~Schlatter,
M.~Schmelling,$^{24}$
M.~Schmitt,
O.~Schneider,
W.~Tejessy,
I.R.~Tomalin,
A.~Venturi,
H.~Wachsmuth,
A.~Wagner
\nopagebreak
\begin{center}
\parbox{15.5cm}{\sl\samepage
European Laboratory for Particle Physics (CERN), 1211 Geneva 23,
Switzerland}
\end{center}\end{sloppypar}
\vspace{2mm}
\begin{sloppypar}
\noindent
Z.~Ajaltouni,
A.~Barr\`{e}s,
C.~Boyer,
A.~Falvard,
P.~Gay,
C~.~Guicheney,
P.~Henrard,
J.~Jousset,
B.~Michel,
S.~Monteil,
J-C.~Montret,
D.~Pallin,
P.~Perret,
F.~Podlyski,
J.~Proriol,
P.~Rosnet,
J.-M.~Rossignol
\nopagebreak
\begin{center}
\parbox{15.5cm}{\sl\samepage
Laboratoire de Physique Corpusculaire, Universit\'e Blaise Pascal,
IN$^{2}$P$^{3}$-CNRS, Clermont-Ferrand, 63177 Aubi\`{e}re, France}
\end{center}\end{sloppypar}
\vspace{2mm}
\begin{sloppypar}
\noindent
T.~Fearnley,
J.B.~Hansen,
J.D.~Hansen,
J.R.~Hansen,
P.H.~Hansen,
B.S.~Nilsson,
B.~Rensch,
A.~W\"a\"an\"anen
\nopagebreak
\begin{center}
\parbox{15.5cm}{\sl\samepage
Niels Bohr Institute, 2100 Copenhagen, Denmark$^{9}$}
\end{center}\end{sloppypar}
\vspace{2mm}
\begin{sloppypar}
\noindent
A.~Kyriakis,
C.~Markou,
E.~Simopoulou,
A.~Vayaki,
K.~Zachariadou
\nopagebreak
\begin{center}
\parbox{15.5cm}{\sl\samepage
Nuclear Research Center Demokritos (NRCD), Athens, Greece}
\end{center}\end{sloppypar}
\vspace{2mm}
\begin{sloppypar}
\noindent
A.~Blondel,
J.C.~Brient,
A.~Roug\'{e},
M.~Rumpf,
A.~Valassi,$^{6}$
H.~Videau$^{21}$
\nopagebreak
\begin{center}
\parbox{15.5cm}{\sl\samepage
Laboratoire de Physique Nucl\'eaire et des Hautes Energies, Ecole
Polytechnique, IN$^{2}$P$^{3}$-CNRS, 91128 Palaiseau Cedex, France}
\end{center}\end{sloppypar}
\vspace{2mm}
\begin{sloppypar}
\noindent
E.~Focardi,$^{21}$
G.~Parrini
\nopagebreak
\begin{center}
\parbox{15.5cm}{\sl\samepage
Dipartimento di Fisica, Universit\`a di Firenze, INFN Sezione di Firenze,
50125 Firenze, Italy}
\end{center}\end{sloppypar}
\vspace{2mm}
\begin{sloppypar}
\noindent
M.~Corden,
C.~Georgiopoulos,
D.E.~Jaffe
\nopagebreak
\begin{center}
\parbox{15.5cm}{\sl\samepage
Supercomputer Computations Research Institute,
Florida State University,
Tallahassee, FL 32306-4052, USA $^{13,14}$}
\end{center}\end{sloppypar}
\vspace{2mm}
\begin{sloppypar}
\noindent
A.~Antonelli,
G.~Bencivenni,
G.~Bologna,$^{4}$
F.~Bossi,
P.~Campana,
G.~Capon,
D.~Casper,
V.~Chiarella,
G.~Felici,
P.~Laurelli,
G.~Mannocchi,$^{5}$
F.~Murtas,
G.P.~Murtas,
L.~Passalacqua,
M.~Pepe-Altarelli
\nopagebreak
\begin{center}
\parbox{15.5cm}{\sl\samepage
Laboratori Nazionali dell'INFN (LNF-INFN), 00044 Frascati, Italy}
\end{center}\end{sloppypar}
\vspace{2mm}
\begin{sloppypar}
\noindent
L.~Curtis,
S.J.~Dorris,
A.W.~Halley,
I.G.~Knowles,
J.G.~Lynch,
V.~O'Shea,
C.~Raine,
P.~Reeves,
J.M.~Scarr,
K.~Smith,
P.~Teixeira-Dias,
A.S.~Thompson,
F.~Thomson,
S.~Thorn,
R.M.~Turnbull
\nopagebreak
\begin{center}
\parbox{15.5cm}{\sl\samepage
Department of Physics and Astronomy, University of Glasgow, Glasgow G12
8QQ,United Kingdom$^{10}$}
\end{center}\end{sloppypar}
\vspace{2mm}
\begin{sloppypar}
\noindent
U.~Becker,
C.~Geweniger,
G.~Graefe,
P.~Hanke,
G.~Hansper,
V.~Hepp,
E.E.~Kluge,
A.~Putzer,
M.~Schmidt,
J.~Sommer,
H.~Stenzel,
K.~Tittel,
S.~Werner,
M.~Wunsch
\begin{center}
\parbox{15.5cm}{\sl\samepage
Institut f\"ur Hochenergiephysik, Universit\"at Heidelberg, 69120
Heidelberg, Fed.\ Rep.\ of Germany$^{16}$}
\end{center}\end{sloppypar}
\vspace{2mm}
\begin{sloppypar}
\noindent
D.~Abbaneo,
R.~Beuselinck,
D.M.~Binnie,
W.~Cameron,
P.J.~Dornan,
P.~Morawitz,
A.~Moutoussi,
J.~Nash,
J.K.~Sedgbeer,
A.M.~Stacey,
M.D.~Williams
\nopagebreak
\begin{center}
\parbox{15.5cm}{\sl\samepage
Department of Physics, Imperial College, London SW7 2BZ,
United Kingdom$^{10}$}
\end{center}\end{sloppypar}
\vspace{2mm}
\begin{sloppypar}
\noindent
G.~Dissertori,
P.~Girtler,
D.~Kuhn,
G.~Rudolph
\nopagebreak
\begin{center}
\parbox{15.5cm}{\sl\samepage
Institut f\"ur Experimentalphysik, Universit\"at Innsbruck, 6020
Innsbruck, Austria$^{18}$}
\end{center}\end{sloppypar}
\vspace{2mm}
\begin{sloppypar}
\noindent
A.P.~Betteridge,
C.K.~Bowdery,
P.~Colrain,
G.~Crawford,
A.J.~Finch,
F.~Foster,
G.~Hughes,
T.~Sloan,
E.P.~Whelan,
M.I.~Williams
\nopagebreak
\begin{center}
\parbox{15.5cm}{\sl\samepage
Department of Physics, University of Lancaster, Lancaster LA1 4YB,
United Kingdom$^{10}$}
\end{center}\end{sloppypar}
\vspace{2mm}
\begin{sloppypar}
\noindent
A.~Galla,
A.M.~Greene,
C.~Hoffmann,
K.~Jacobs,
K.~Kleinknecht,
G.~Quast,
B.~Renk,
E.~Rohne,
H.-G.~Sander,
P.~van~Gemmeren
C.~Zeitnitz
\nopagebreak
\begin{center}
\parbox{15.5cm}{\sl\samepage
Institut f\"ur Physik, Universit\"at Mainz, 55099 Mainz, Fed.\ Rep.\
of Germany$^{16}$}
\end{center}\end{sloppypar}
\vspace{2mm}
\begin{sloppypar}
\noindent
J.J.~Aubert,$^{21}$
A.M.~Bencheikh,
C.~Benchouk,
A.~Bonissent,
G.~Bujosa,
D.~Calvet,
J.~Carr,
C.~Diaconu,
N.~Konstantinidis,
P.~Payre,
D.~Rousseau,
M.~Talby,
A.~Sadouki,
M.~Thulasidas,
A.~Tilquin,
K.~Trabelsi
\nopagebreak
\begin{center}
\parbox{15.5cm}{\sl\samepage
Centre de Physique des Particules, Facult\'e des Sciences de Luminy,
IN$^{2}$P$^{3}$-CNRS, 13288 Marseille, France}
\end{center}\end{sloppypar}
\vspace{2mm}
\begin{sloppypar}
\noindent
M.~Aleppo, 
F.~Ragusa$^{21}$
\nopagebreak
\begin{center}
\parbox{15.5cm}{\sl\samepage
Dipartimento di Fisica, Universit\'a di Milano e INFN Sezione di
Milano, 20133 Milano, Italy.}
\end{center}\end{sloppypar}
\vspace{2mm}
\begin{sloppypar}
\noindent
C.~Bauer,
R.~Berlich,
W.~Blum,
V.~B\"uscher,
H.~Dietl,
F.~Dydak,$^{21}$
G.~Ganis,
C.~Gotzhein,
H.~Kroha,
G.~L\"utjens,
G.~Lutz,
W.~M\"anner,
H.-G.~Moser,
R.~Richter,
A.~Rosado-Schlosser,
S.~Schael,
R.~Settles,
H.~Seywerd,
R.~St.~Denis,
H.~Stenzel,
W.~Wiedenmann,
G.~Wolf
\nopagebreak
\begin{center}
\parbox{15.5cm}{\sl\samepage
Max-Planck-Institut f\"ur Physik, Werner-Heisenberg-Institut,
80805 M\"unchen, Fed.\ Rep.\ of Germany\footnotemark[16]}
\end{center}\end{sloppypar}
\vspace{2mm}
\begin{sloppypar}
\noindent
J.~Boucrot,
O.~Callot,
A.~Cordier,
M.~Davier,
L.~Duflot,
J.-F.~Grivaz,
Ph.~Heusse,
A.~H\"ocker,
A.~Jacholkowska,
M.~Jacquet,
D.W.~Kim\rlap,$^{19}$
F.~Le~Diberder,
J.~Lefran\c{c}ois,
A.-M.~Lutz,
I.~Nikolic,
H.J.~Park,$^{19}$
M.-H.~Schune,
S.~Simion,
J.-J.~Veillet,
I.~Videau,
D.~Zerwas
\nopagebreak
\begin{center}
\parbox{15.5cm}{\sl\samepage
Laboratoire de l'Acc\'el\'erateur Lin\'eaire, Universit\'e de Paris-Sud,
IN$^{2}$P$^{3}$-CNRS, 91405 Orsay Cedex, France}
\end{center}\end{sloppypar}
\vspace{2mm}
\begin{sloppypar}
\noindent
\samepage
P.~Azzurri,
G.~Bagliesi,
G.~Batignani,
S.~Bettarini,
C.~Bozzi,
G.~Calderini,
M.~Carpinelli,
M.A.~Ciocci,
V.~Ciulli,
R.~Dell'Orso,
R.~Fantechi,
I.~Ferrante,
A.~Giassi,
A.~Gregorio,
F.~Ligabue,
A.~Lusiani,
P.S.~Marrocchesi,
A.~Messineo,
F.~Palla,
G.~Rizzo,
G.~Sanguinetti,
A.~Sciab\`a,
P.~Spagnolo,
J.~Steinberger,
R.~Tenchini,
G.~Tonelli,$^{26}$
C.~Vannini,
P.G.~Verdini,
J.~Walsh
\samepage
\begin{center}
\parbox{15.5cm}{\sl\samepage
Dipartimento di Fisica dell'Universit\`a, INFN Sezione di Pisa,
e Scuola Normale Superiore, 56010 Pisa, Italy}
\end{center}\end{sloppypar}
\vspace{2mm}
\begin{sloppypar}
\noindent
G.A.~Blair,
L.M.~Bryant,
F.~Cerutti,
J.T.~Chambers,
Y.~Gao,
M.G.~Green,
T.~Medcalf,
P.~Perrodo,
J.A.~Strong,
J.H.~von~Wimmersperg-Toeller
\nopagebreak
\begin{center}
\parbox{15.5cm}{\sl\samepage
Department of Physics, Royal Holloway \& Bedford New College,
University of London, Surrey TW20 OEX, United Kingdom$^{10}$}
\end{center}\end{sloppypar}
\vspace{2mm}
\begin{sloppypar}
\noindent
D.R.~Botterill,
R.W.~Clifft,
T.R.~Edgecock,
S.~Haywood,
P.~Maley,
P.R.~Norton,
J.C.~Thompson,
A.E.~Wright
\nopagebreak
\begin{center}
\parbox{15.5cm}{\sl\samepage
Particle Physics Dept., Rutherford Appleton Laboratory,
Chilton, Didcot, Oxon OX11 OQX, United Kingdom$^{10}$}
\end{center}\end{sloppypar}
\vspace{2mm}
\begin{sloppypar}
\noindent
B.~Bloch-Devaux,
P.~Colas,
S.~Emery,
W.~Kozanecki,
E.~Lan\c{c}on,
M.C.~Lemaire,
E.~Locci,
B.~Marx,
P.~Perez,
J.~Rander,
J.-F.~Renardy,
A.~Roussarie,
J.-P.~Schuller,
J.~Schwindling,
A.~Trabelsi,
B.~Vallage
\nopagebreak
\begin{center}
\parbox{15.5cm}{\sl\samepage
CEA, DAPNIA/Service de Physique des Particules,
CE-Saclay, 91191 Gif-sur-Yvette Cedex, France$^{17}$}
\end{center}\end{sloppypar}
\vspace{2mm}
\begin{sloppypar}
\noindent
S.N.~Black,
J.H.~Dann,
R.P.~Johnson,
H.Y.~Kim,
A.M.~Litke,
M.A. McNeil,
G.~Taylor
\nopagebreak
\begin{center}
\parbox{15.5cm}{\sl\samepage
Institute for Particle Physics, University of California at
Santa Cruz, Santa Cruz, CA 95064, USA$^{22}$}
\end{center}\end{sloppypar}
\vspace{2mm}
\begin{sloppypar}
\noindent
C.N.~Booth,
R.~Boswell,
C.A.J.~Brew,
S.~Cartwright,
F.~Combley,
A.~Koksal,
M.~Letho,
W.M.~Newton,
J.~Reeve,
L.F.~Thompson
\nopagebreak
\begin{center}
\parbox{15.5cm}{\sl\samepage
Department of Physics, University of Sheffield, Sheffield S3 7RH,
United Kingdom$^{10}$}
\end{center}\end{sloppypar}
\vspace{2mm}
\begin{sloppypar}
\noindent
A.~B\"ohrer,
S.~Brandt,
G.~Cowan,
C.~Grupen,
P.~Saraiva,
L.~Smolik,
F.~Stephan,
\nopagebreak
\begin{center}
\parbox{15.5cm}{\sl\samepage
Fachbereich Physik, Universit\"at Siegen, 57068 Siegen,
 Fed.\ Rep.\ of Germany$^{16}$}
\end{center}\end{sloppypar}
\vspace{2mm}
\begin{sloppypar}
\noindent
M.~Apollonio,
L.~Bosisio,
R.~Della~Marina,
G.~Giannini,
B.~Gobbo,
G.~Musolino
\nopagebreak
\begin{center}
\parbox{15.5cm}{\sl\samepage
Dipartimento di Fisica, Universit\`a di Trieste e INFN Sezione di Trieste,
34127 Trieste, Italy}
\end{center}\end{sloppypar}
\vspace{2mm}
\begin{sloppypar}
\noindent
J.~Putz,
J.~Rothberg,
S.~Wasserbaech,
R.W.~Williams
\nopagebreak
\begin{center}
\parbox{15.5cm}{\sl\samepage
Experimental Elementary Particle Physics, University of Washington, WA 98195
Seattle, U.S.A.}
\end{center}\end{sloppypar}
\vspace{2mm}
\begin{sloppypar}
\noindent
S.R.~Armstrong,
P.~Elmer,
Z.~Feng,$^{12}$
D.P.S.~Ferguson,
Y.S.~Gao,$^{23}$
S.~Gonz\'{a}lez,
J.~Grahl,
T.C.~Greening,
O.J.~Hayes,
H.~Hu,
P.A.~McNamara III,
J.M.~Nachtman,
W.~Orejudos,
Y.B.~Pan,
Y.~Saadi,
I.J.~Scott,
A.M.~Walsh,$^{27}$
Sau~Lan~Wu,
X.~Wu,
J.M.~Yamartino,
M.~Zheng,
G.~Zobernig
\nopagebreak
\begin{center}
\parbox{15.5cm}{\sl\samepage
Department of Physics, University of Wisconsin, Madison, WI 53706,
USA$^{11}$}
\end{center}\end{sloppypar}
}
\footnotetext[1]{Now at DESY, Hamburg, Germany.}
\footnotetext[2]{Supported by Direcci\'on General de Investigaci\'on
Cient\'ifica y T\'ecnica, Spain.}
\footnotetext[3]{Now at Dipartimento di Fisica, Universit\`{a} di Lecce,
73100 Lecce, Italy.}
\footnotetext[4]{Also Istituto di Fisica Generale, Universit\`{a} di
Torino, Torino, Italy.}
\footnotetext[5]{Also Istituto di Cosmo-Geofisica del C.N.R., Torino,
Italy.}
\footnotetext[6]{Supported by the Commission of the European Communities,
contract ERBCHBICT941234.}
\footnotetext[7]{Supported by CICYT, Spain.}
\footnotetext[8]{Supported by the National Science Foundation of China.}
\footnotetext[9]{Supported by the Danish Natural Science Research Council.}
\footnotetext[10]{Supported by the UK Particle Physics and Astronomy Research
Council.}
\footnotetext[11]{Supported by the US Department of Energy, grant
DE-FG0295-ER40896.}
\footnotetext[12]{Now at The Johns Hopkins University, Baltimore, MD 21218, U.S.A.}
\footnotetext[13]{Supported by the US Department of Energy, contract
DE-FG05-92ER40742.}
\footnotetext[14]{Supported by the US Department of Energy, contract
DE-FC05-85ER250000.}
\footnotetext[15]{Permanent address: Universitat de Barcelona, 08208 Barcelona,
Spain.}
\footnotetext[16]{Supported by the Bundesministerium f\"ur Forschung und
Technologie, Fed. Rep. of Germany.}
\footnotetext[17]{Supported by the Direction des Sciences de la
Mati\`ere, C.E.A.}
\footnotetext[18]{Supported by Fonds zur F\"orderung der wissenschaftlichen
Forschung, Austria.}
\footnotetext[19]{Permanent address: Kangnung National University,
Kangnung, Korea.}
\footnotetext[20]{Now at CERN, 1211 Geneva 23, Switzerland.}
\footnotetext[21]{Also at CERN, 1211 Geneva 23,
Switzerland.}
\footnotetext[22]{Supported by the US Department of Energy,
grant DE-FG03-92ER40689.}
\footnotetext[23]{Now at Harvard University, Cambridge, MA 02138, U.S.A.}
\footnotetext[24]{Now at Max-Plank-Instit\"ut f\"ur Kernphysik, Heidelberg,
Germany.}
\footnotetext[25]{Now at Dragon Systems, Newton, MA 02160, U.S.A.}
\footnotetext[26]{Also at Istituto di Matematica e Fisica,
Universit\`a di Sassari, Sassari, Italy.}
\footnotetext[27]{Now at Rutgers University, Piscataway, NJ 08855-0849, U.S.A.}
%
%
\setlength{\parskip}{\saveparskip}
\setlength{\textheight}{\savetextheight}
\setlength{\topmargin}{\savetopmargin}
\setlength{\textwidth}{\savetextwidth}
\setlength{\oddsidemargin}{\saveoddsidemargin}
\setlength{\topsep}{\savetopsep}
\normalsize
\newpage
\pagestyle{plain}
\setcounter{page}{1}
\pretolerance=1000
\section{Introduction}
\setcounter{page}{1}
\par
The lightest supersymmetric particle (LSP) commonly is assumed to be the
lightest neutralino, $\chi$.  This assumption follows naturally if R-parity,
which distinguishes supersymmetric and ordinary particles, is conserved, in
which case the LSP must be stable~\cite{susy}.  Cosmological
considerations~\cite{bang} rule out an LSP with charge or color, leaving only
the lightest neutralino and the sneutrinos as possible LSP's.
While a lower limit on the sneutrino mass of~$43~\GeVcsq$ can be derived from
the invisible~Z width, previous analyses have not excluded a massless
neutralino, hence the current focus on neutralinos.  However, the improved
limits on $\Mchi$ presented here do not require sneutrinos to be heavier than
neutralinos.  Since the lightest neutralino, if stable, could constitute a
significant fraction of the dark matter of the universe~\cite{weinberg},
bounds on its mass from high energy physics experiments are quite
relevant~\cite{olive,cosmo}.
\par
Searches for charginos, neutralinos and sleptons have been carried out
using data taken with the \ALEPH~detector~\cite{alephdet} at
\lepa~($\sqrt{s}\sim M_{\mathrm{Z}}$) and more recently at
\lepb~($\sqrt{s} = 130-136~\GeV$). The negative results of those searches
are reported in Refs.~\cite{gup,susy133}, where details of the search
techniques and definitions of terms can be found.  It was assumed that the
lightest neutralino is stable and escapes detection, causing an apparent
missing energy in the event.
\par
The results of \lepa~and \lepb~chargino and neutralino searches together place
significant bounds on the mass of the lightest neutralino, as described in
this article. While in each case open regions in parameter space exist
which allow a very light, even massless, neutralino, it turns out that these
regions are not in common. For large sneutrino masses, a massless neutralino
allowed by the \lepa~analysis is excluded by \lepb, and vice-versa.
\par
The combination of \lepa~and \lepb~results is fruitful in the scalar lepton
sector also.  The limit on the sneutrino mass coming from the Z~width and
the limits from the direct search for sleptons taken together exclude a
significant region in parameter space when their masses are linked according
to the ideas of SUSY-based grand unified theories.  This exclusion can be
used to constrain the mass of the neutralino in a way which complements the
limits from the charginos and neutralinos.
\par
The non-observation of Higgs bosons at \lepa\ can also be used to restrict
neutralino masses in a highly constrained SUSY GUT called ``minimal
supergravity.'' In this theory, the derivation of electroweak symmetry breaking
through radiative corrections induced by the top quark Yukawa coupling allows
a reduction in the number of SUSY parameters, leading ultimately to a stronger
limit on $\Mchi$.
\section{Analysis}
\par
In the Minimal Supersymmetric Standard Model (MSSM), chargino ($\cha$) and
neutralino ($\chi$, $\chip$, $\chipp$, $\chippp$) masses depend on the
parameters $M_1$, $M_2$, $\mu$, and $\tb$. The gauge fermion masses at the
electroweak scale are denoted by $M_1$ and $M_2$. If they are equal at the
GUT scale, then $M_1={{5}\over{3}}\tan^2\theta_{\mathrm W}\,M_2$, which is
assumed here.  The independent parameter $\mu$ represents the Higgsino mass,
and $\tb$ is the ratio of Higgs doublet expectation values.  Since the top
quark is much heavier than the bottom quark, one expects $\tb>1$. Furthermore,
if the top quark Yukawa coupling remains perturbative up to the GUT scale, then
$\tb>1.2$~\cite{mintb}.  The restriction $\tb\ge 1$ is imposed in this
analysis, but it is worth noting that the chargino and neutralino masses
and couplings are symmetric under the transformation $\tb\goto\cot\beta$.
Tree-level relations for all masses are used, as radiative corrections
for these particles are small~\cite{treemass}.
\par
\begin{figure}
\begin{center}
\mbox{\epsfig{file=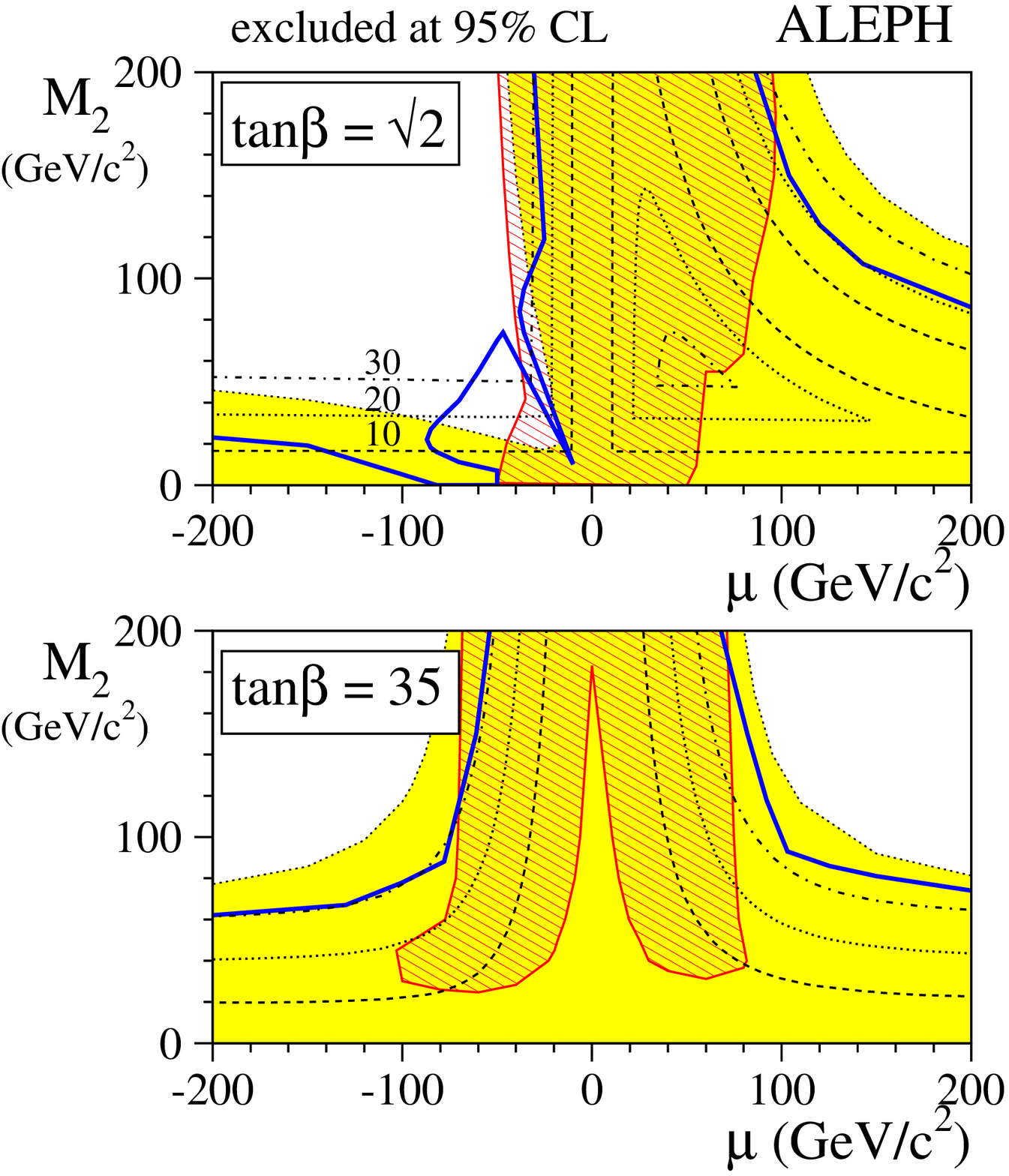,height=16.5cm%
,bbllx=1mm,bblly=30mm,bburx=160mm,bbury=220mm}}
\end{center}
\caption[.]{\em
Excluded regions in the $(\mu,M_2$) plane, for $\tb=\sqrt{2}$
and~$35$, computed for $\Msnu = 500~\GeVcsq$.
The shaded (hatched) region is excluded by the chargino (neutralino)
search at~\lepb.  The heavy black curve shows the region excluded at \lepa.
Finally, curves of constant neutralino mass ($\Mchi=10$, $20$, and
$30~\GeVcsq$) are shown.
\label{muM2detail} }
\end{figure}

The functional dependence of the chargino and neutralino masses on the SUSY
parameters is different. Consequently, for any given chargino mass,
there is a smallest neutralino mass, possibly zero.  In this analysis,
bounds in the $(\mu,M_2)$ plane for a series of $\tb$ and sneutrino
mass values are derived using the \ALEPH~analyses to obtain experimental
efficiencies and to investigate special cases.  The {\sc SUSYGEN}
program~\cite{susygen} was used to generate chargino, neutralino, and slepton
events in relevant regions of the SUSY parameter space.  For each $\tb$ and
sneutrino mass, the point in parameter space outside all regions excluded
at~95\% CL giving the lowest neutralino mass is found.  Fig.~\ref{muM2detail}
allows a comparison of the lines of constant $\Mchi$ mass with these
experimental bounds in the $(\mu,M_2)$ plane.
\par
The slepton and sneutrino masses play a key role in the production and
decay of charginos and neutralinos.  The destructive interference between
$s$-channel and $t$-channel diagrams reduces the chargino cross section
when the electron-sneutrino is lighter than~$70~\GeVcsq$.
Constructive interference increases associated neutralino production when
the selectron is light.  The three-body decay of charginos and
neutralinos proceeds mainly through intermediate W~and Z~exchange when
all sneutrinos and sleptons are heavy.  When they are light enough, however,
two-body decays dominate, with radical consequences. In particular, the
final state $\chi^+\goto\ell^+\snu$ will not be selected if the mass
difference $M_{\chi^+}-M_{\snu}$ is smaller than about $3~\GeVcsq$, leaving
an experimental `blind spot' in the exclusion of charginos heavier than
$45~\GeVcsq$, the limit inferred from the~Z width.  Similarly, the dominance
of the decay $\chip\goto\nu\snu$ when $\Msnu<\Mchip$ leads to a complete loss
of experimental acceptance for the $\chi\chip$ final state.  The cases of
heavy and light sneutrinos and sleptons are discussed in detail in the
following sections.
\subsection{Heavy Sneutrinos and Sleptons}
\label{a1}
\par
Large sneutrino and slepton masses were assumed when deriving the excluded
regions in the $(\mu,M_2)$ plane depicted in Fig.~5 of Ref.~\cite{susy133},
a detail of which is shown in Fig.~\ref{muM2detail}. In this case the
influence of the $t$-channel diagram in the production cross section is small,
and the decay branching fractions of charginos and neutralinos are
approximately the same as those of~W and~Z bosons.
\par
The chargino search at \lepb~provides the most stringent bound on $\Mchi$
as a function of $\mu$ for $\tb\appgt 2$.
When $\tb\appgt 10$, the lowest value of $\Mchi$
is found deep in the gaugino region ($\mu\approx -4$~TeV/c$^2$).  Although
such regions are not favoured theoretically, the difference in the $\Mchi$
bound between $\mu=-4~{\mathrm{TeV}}/c^2$ and~$-500~\GeVcsq$ is less than
$0.2~\GeVcsq$, so theoretical prejudice on the range of~$\mu$ need not be
taken into account.  As $\tb$ approaches~2, the minimum neutralino mass is
found for $\mu$ near~$-70~\GeVcsq$.
\par
\begin{figure}
\begin{center}
\mbox{\epsfig{file=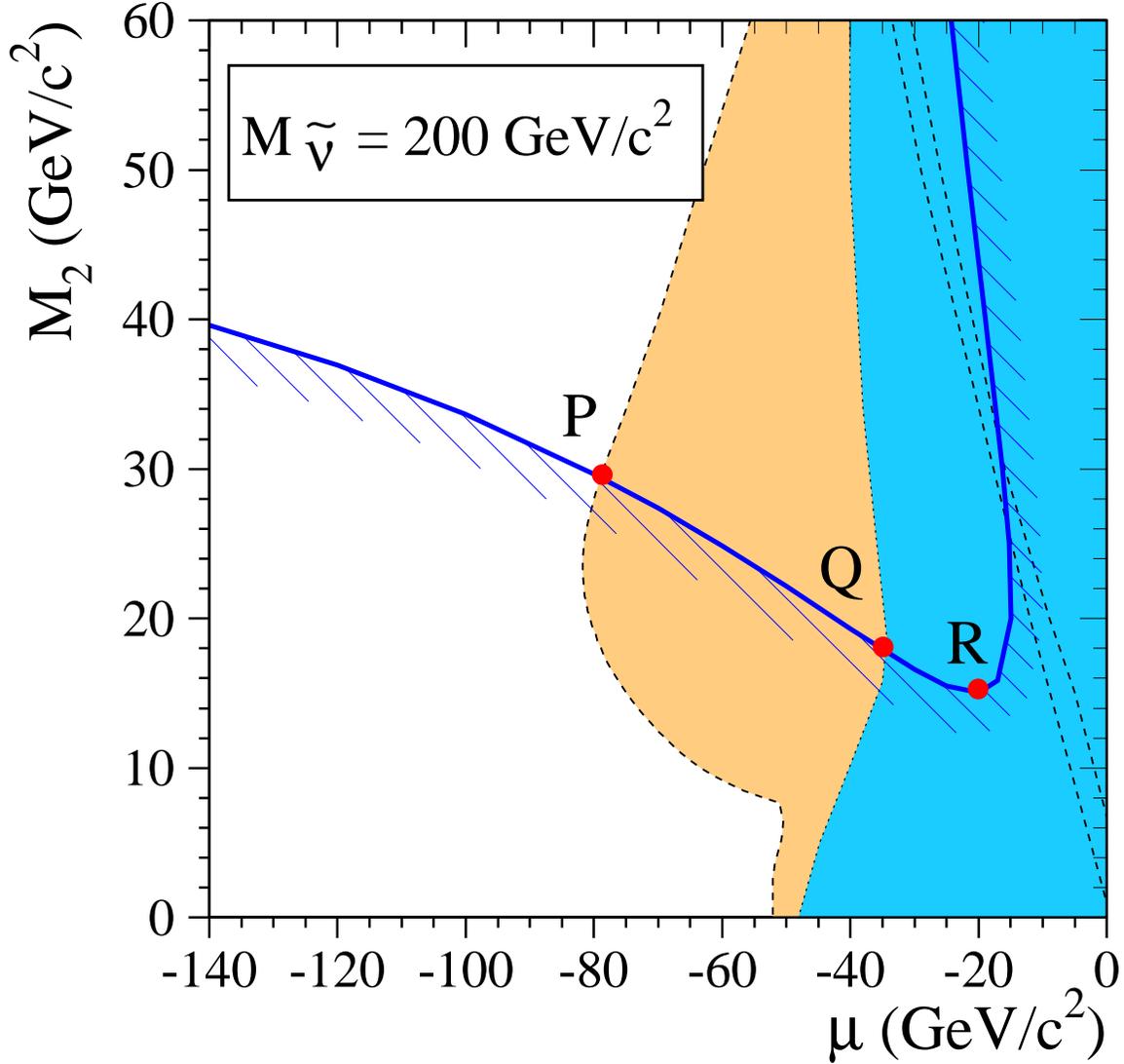,height=16.5cm%
,bbllx=20mm,bblly=50mm,bburx=155mm,bbury=225mm}}
\end{center}
\caption[.]{\em
Close-up view of the limit contours coming from the
chargino search at \lepb~(heavy solid curve),
from the neutralino search at \lepb~(dark shaded region),
and the direct search for neutralinos at \lepa~(light shaded region),
for $\tan\beta=\sqrt{2}$ and $\Msnu = 200~\GeVcsq$.
(Additional exclusion contours from the Z~width are not shown as
they do not play any role in the bound on $\Mchi$ for this value
of $\tb$.)
Point~{\sc Q} is the intersection of the chargino
and neutralino searches from \lepb~which gives a slightly better limit
on $\Mchi$ than the chargino limit alone (point~{\sc R}).
Point~{\sc P} is the intersection of the chargino and \lepa~neutralino
searches, which sets the best limit on $\Mchi$.
The thin strip not excluded by \lepa~is indicated by the dashed curves;
it is excluded by neutralino searches at \lepb.
\label{closeup_large_m0} }
\end{figure}

When $\tb$ falls below~$2$, the neutralino search at \lepb~excludes additional
regions of parameter space, as can be seen clearly for $\tb=\sqrt{2}$ in
Fig.~\ref{closeup_large_m0}. {\sc R} marks the point at which the chargino
bound places a lower limit on $\Mchi$; this limit is improved at
point~{\sc Q} by the \lepb\ neutralino bound.  The improvement is significant
for $1.1<\tb<1.5$.  For example, when $\tb=1.35$, the chargino limits alone
do not exclude a massless neutralino, but the neutralino results place the
bound at $\Mchi>10~\GeVcsq$.
\par
\begin{figure}
\begin{center}
\mbox{\epsfig{file=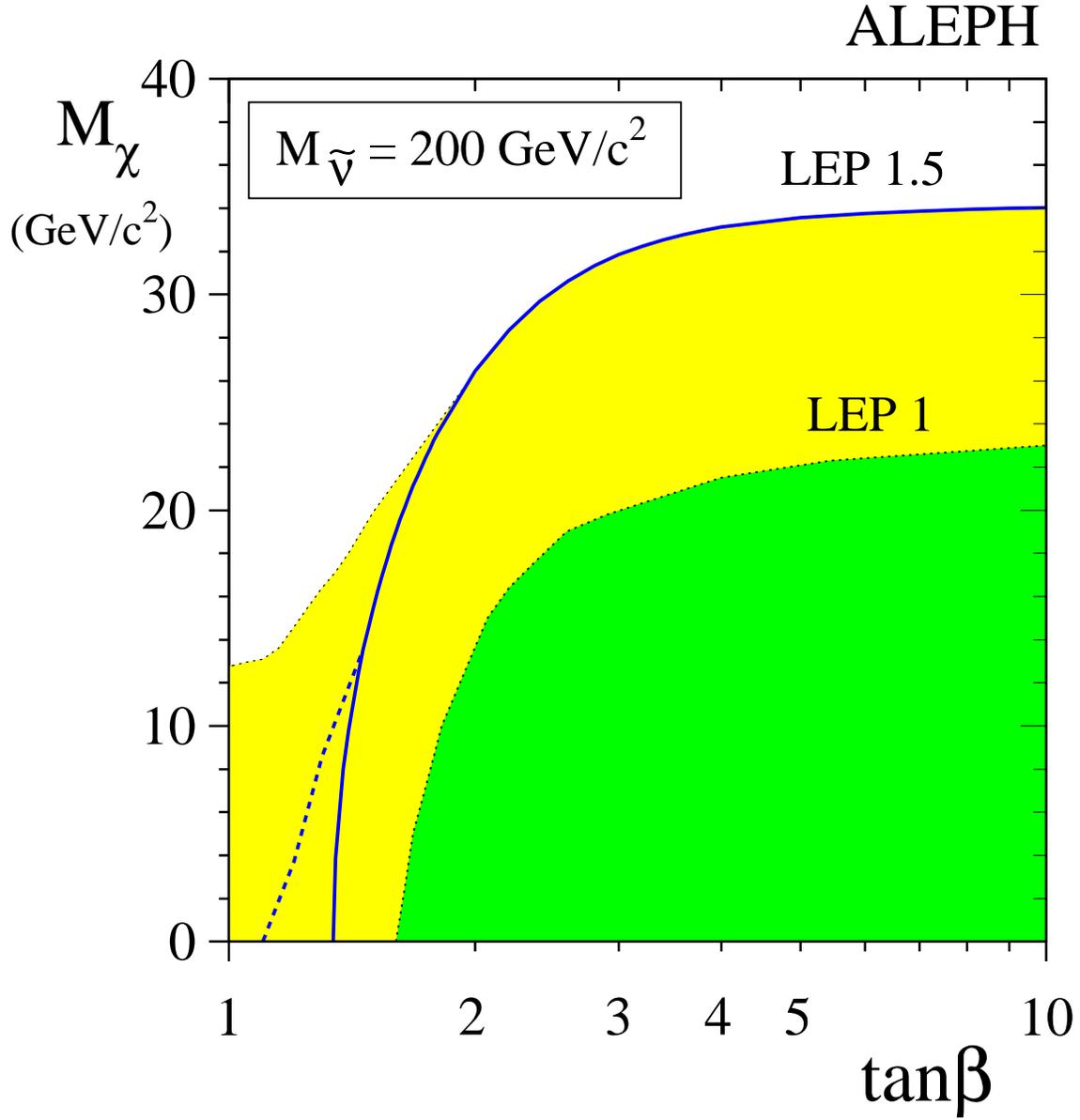,height=16cm%
,bbllx=1mm,bblly=55mm,bburx=165mm,bbury=225mm}}
\end{center}
\caption[.]{\em
Lower limit on the mass of the lightest neutralino as a
function of $\tb$, for $\Msnu=200~\GeVcsq$.
The previous limit from \lepa~is shown as the dark area.
The solid curve shows the limit obtained by the \lepb~chargino search alone,
with the small extension coming from the \lepb~neutralino search shown as
a dashed curve.  The combined result is given by the light shaded area.
\label{mchi_limit} }
\end{figure}

The reach of the \lepb~neutralino bounds is limited to the higgsino region
(typically $|\mu|\applt 55~\GeVcsq$), and does not suffice to exclude massless
neutralinos for $\tb < 1.1$.  Fortuitously, the direct search for neutralinos
at \lepa~covers the region not covered by \lepb. The production of $\chi\chip$
and $\chip\chip$ in~Z decays would be large, so this region is excluded by the
direct search for neutralinos at \lepa~\cite{gup}, as shown in
Fig.~\ref{closeup_large_m0}, where the point~{\sc P} marks the limit obtained
from the \lepb\ and \lepa\ excluded regions taken together. The results from
\lepa~start to play a role for $\tb\applt 2$, and, taken together with the
chargino and neutralino results from \lepb, exclude for any $\tb$ a lightest
neutralino with mass less than~$12.8~\GeVcsq$ at 95\% CL.  This bound is shown
in Fig.~\ref{mchi_limit}, along with the previous bound from \lepa, and the
intermediate results from the \lepb~chargino and neutralino searches.
It does not depend on any specific relations among sneutrino and slepton
masses provided they are all greater than approximately~$100~\GeVcsq$.
\par
A minor exception to this bound on $\Mchi$ deserves some comment.  A close-up
view of the relevant excluded regions in the $(\mu,M_2)$ plane is shown in
Fig.~\ref{closeup_101}, for $\tb=1.01$. A small hole not covered by any of
these analyses is indicated near $\mu = -30~\GeVcsq$ and $M_2 = 3~\GeVcsq$.
At centre-of-mass energies reached by \lep\ so far, chargino production in
this region is not possible. Furthermore, only the couplings of~$\chip\chipp$
and~$\chip\chippp$ to the~Z are non-vanishing.  The latter, however, is
kinematically forbidden, leaving $\chip\chipp$ alone as a possible signal
channel.  The cross section is about $0.7$~pb, allowing exclusion for roughly
twice the luminosity recorded by \ALEPH. One might expect, therefore, that a
combination of the neutralino searches performed by the \lep~Collaborations
would show that hole to be covered. The analysis of Ref.~\cite{gup} was
performed using fewer than $2\times 10^5$ Z decays.  An update using the full
data sample would reduce the size of this hole but not eliminate it
completely. It exists only for $\tb < 1.02$ and $\Msnu>80~\GeVcsq$, and
hereafter will be ignored. The bound $\Mchi > 12.8~\GeVcsq$ is given by the
point~{\sc P} in Fig.~\ref{closeup_101}.
\par
The \lepa~results alone cannot rule out a massless neutralino for low
$\tb$ because the chargino mass limit does not overlap the limit from the
neutralino search sufficiently.  The improved bounds presented in
Fig.~\ref{mchi_limit} do not result simply from the higher centre-of-mass
energies at \lepb, but rather from a particular juxtaposition of excluded
regions.  In this light it is worth noting that $\chi\chip$ production
excluded by the \lepb~neutralino search covers a narrow strip along
$M_2\approx -2|\mu|$ not ruled out by \lepa~searches
(Fig.~\ref{closeup_large_m0}) due to a vanishing coupling of the~Z to
$\chi\chip$.
\par
The importance of the neutralino searches in the small $|\mu|$, small $M_2$
regime is evident.  For part of this region the final state can contain one or
two $\chip\,$'s, which decay to $\chi\gamma$ with a large branching
fraction~\cite{radchi}.  Even though the $\chip$ and $\chi$ can be very
light, they are boosted and the photon is energetic.  The results presented
here were made insensitive to the details of these radiative decays by
augmenting the selection of Ref.~\cite{susy133} with one which requires an
energetic, isolated photon, and with the search for acoplanar photon pairs
reported in Ref.~\cite{multiphotons}.  Details are given in the Appendix.
%
\subsection{Light Sneutrinos and Sleptons}
\label{a2}
\par
The impact of light sneutrinos and sleptons in the production and decay of
charginos and neutralinos can be investigated taking into account experimental
bounds.  The most general case is too complicated to be useful, however, so
the available information is best organized by making some additional,
theoretically well-motivated assumptions.
\par
In supergravity-inspired GUTs all SUSY scalar particles have a common mass
$m_0$ and all SUSY fermions have a common mass $\mhalf$ at the GUT
scale~\cite{sugra}. After evolution from the GUT to the electroweak scale
according to the renormalization group equations~\cite{rge}, the following
formulae for slepton and sneutrino masses apply:
\begin{eqnarray}
\label{m_scalers}
  m^2_{\tilde{\ell}_R} &=&
        m^2_0 + 0.15\,m^2_{1/2} - \ssqtw M^2_{\mathrm{Z}} \cos 2\beta \cr
  m^2_{\tilde{\ell}_L} &=&
        m^2_0 + 0.52\,m^2_{1/2} -
            \onehalf
           (1-2\ssqtw) M^2_{\mathrm{Z}} \cos 2\beta \cr
  m^2_{\tilde{\nu}} &=&
        m^2_0 + 0.52\,m^2_{1/2} +
            \onehalf
           M^2_{\mathrm{Z}} \cos 2\beta .
\end{eqnarray}
According to these assumptions, all charged left-sleptons have the same
mass, and all right-sleptons have the same mass, always smaller than that of
the left-sleptons.  The theoretical parameters $\mhalf$ and $M_2$ are
interchangeable:
\begin{equation}
\label{m2_m12}
    M_2 = 0.82\, \mhalf
\end{equation}
at the electroweak scale.  Consequently, the gaugino and the slepton masses
are linked.
\par
The sensitivity of the $\Mchi$ limit (Fig.~\ref{mchi_limit}) to light
sneutrino and selectron masses can be investigated coherently by use of
these relations.  The results of the previous subsection correspond to
$m_0 = 200~\GeVcsq$.  For smaller values of $m_0$, the changes in the
production and decay of charginos and neutralinos follow a pattern which
can be illustrated by the following example.
\par
Relevant excluded regions in a limited portion of the $(\mu,M_2)$ plane are
shown in Figs.~\ref{closeup_large_m0},~\ref{closeup_medium_m0},
and~\ref{closeup_small_m0}, for $\tb=\sqrt{2}$ and decreasing values of $m_0$.
As discussed previously, the sneutrinos and sleptons have little influence
for $m_0=200~\GeVcsq$ (Fig.~\ref{closeup_large_m0}), and the lowest allowed
value for $\Mchi$ is given by the intersection point~{\sc P} of the
\lepb~chargino curve and the \lepa~direct search for neutralinos.  For
$m_0=68~\GeVcsq$ however (Fig.~\ref{closeup_medium_m0}), the limit is weaker
due to the effects described in the following. If the electron-sneutrino and
selectron had no influence on the production or decay of charginos and
neutralinos, then point~{\sc P} would mark the limit on $\Mchi$.  Taking into
account the decrease of the chargino cross section due to the low
electron-sneutrino mass ($\Msnu\approx 62~\GeVcsq$), the limit falls to
point~{\sc P}$^\prime$.  At that point, however, the two-body decays
$\chi^\pm\goto\ell^\pm\snu$ and $\chip\goto\nu\snu$ dominate, and
since the sneutrino mass is close to the chargino and neutralino masses
($\Mcha\approx 64~\GeVcsq$ and $\Mchip\approx 65~\GeVcsq$), the point is not
excluded.  The actual experimental limit is given by
point~{\sc P}$^{\prime\prime}$, at which $\Mchi=9.2~\GeVcsq$.
This erosion of the excluded region from~{\sc P$^\prime$}
to~{\sc P}$^{\prime\prime}$ is a direct consequence of the sharp transition
from three-body to two-body decays mentioned above.  The actual `blind spot'
in the exclusion of charginos at \lepb~is shown in Fig.~\ref{cha_2_3}.
Although a substantial region of two-body decays is excluded, it does not
strengthen the lower limit on~$\Mchi$.
\par
Given the importance of sneutrinos and sleptons on the~$\Mchi$ limit, it is
worth examining the bounds on their masses.  The analysis of
Ref.~\cite{susy133} is employed to derive bounds on the production of
charged sleptons.  Masses are calculated as a function of $m_0$, $M_2$,
and $\tb$ according to Eqs.~(\ref{m_scalers}) and~(\ref{m2_m12}).
Production cross sections for selectrons and decay branching ratios
for left-sleptons depend on couplings to and masses of the gauginos, so~$\mu$
plays an implicit role.  Fixing $\tb$, limits in the $(M_2,m_0)$ plane can be
derived, taking the values for~$\mu$ which give the most conservative limit.
Only those ranges of $\mu$ not already excluded by chargino and neutralino
bounds are considered.  In practice, the weakest limit on $M_2$ for a given
$m_0$ is obtained for $-300\applt\mu\applt-50~\GeVcsq$.
\par
Significant mixing between $\tilde{\tau}_{\mathrm{R}}$ and
$\tilde{\tau}_{\mathrm{L}}$ can occur for large $\tb$, resulting in a light
mass eigenstate $\tilde{\tau}_1$.  Although the production rate for the
light $\tilde{\tau}_1$ would be larger than for the flavour eigenstate
$\tilde{\tau}_{\mathrm{R}}$, this mixing between left and right-sleptons has
been ignored.
\par
Recent precise measurements of the~Z line shape~\cite{elw95} imply a bound
on the sneutrino mass: $\Msnu > 43~\GeVcsq$ at 95\% CL. Generally speaking,
the sleptons exclude a larger region in the $(M_2,m_0)$ plane than
the sneutrino mass bound when $\tb\applt 2$.  However, due to the different
sign in the coefficients of the $\cos 2\beta$ terms (Eq.~\ref{m_scalers}),
the latter bound is stronger for $\tb > 2$.
\par
Fig.~\ref{slep_m0_m12} shows the regions excluded in the $(M_2,m_0)$ plane,
for $\tb = 1$, $\sqrt{2}$, $2$, and $35$, computed for $\mu<0$. For $\tb=35$
the exclusion comes solely from the sneutrino mass, while for $\tb=2$, both
sneutrinos and sleptons play a role (the transition occurs at
$M_2 = 62~\GeVcsq$).  Sleptons dominate the limit for $\tb = 1$ and~$\sqrt{2}$.
The unusual shape for $M_2 > 60~\GeVcsq$ results from a production
cross section which increases as $m_0$ approaches zero, and an experimental
acceptance which decreases due to small differences between slepton and
neutralino masses.  For $\tb=1$, in fact, a significant region is disallowed
theoretically by the requirement that the neutralino be lighter than the
charged sleptons, as indicated by the dashed line.  Slightly above this
theoretical limit the experimental acceptance is zero because the lepton
momenta are too low.  At $M_2\sim 44~\GeVcsq$, however, the
sneutrino limit coming from the~Z width applies, as shown.
\par
In the large $M_2$ region the mass difference
$M_{\tilde{\ell}_L}-M_{\tilde{\ell}_R}$ is large, and light selectrons,
smuons, and staus all contribute to the limit.  The associated production of
$\tilde{\mathrm{e}}_L \tilde{\mathrm{e}}_R$ contributes also,
with both left- and right-sleptons decaying purely to $\ell\chi$.
In contrast, the bound for small $M_2$ comes mainly from
$\tilde{\mathrm{e}}_L$ and $\tilde{\mathrm{e}}_R$ pairs
which are nearly mass degenerate, and the contribution from
$\tilde{\mathrm{e}}_L\tilde{\mathrm{e}}_R$ is negligible.  Although the
left-sleptons develop a large branching fraction ($\appgt 60$\%) to
$\chi^\pm\nu$ for large negative $\mu$, the acceptance is only a few
percent lower than for the single-prong $\ell\chi$ final state, because the
event selection includes topologies with leptons and hadrons as well
as acoplanar pairs~\cite{susy133}.
\par
The slepton and sneutrino bounds on $M_2$ as a function of $m_0$ supplement
the limits on $\Mchi$ derived above.  An example of the impact of slepton
bounds in the $(\mu,M_2)$ plane is shown in Fig.~\ref{closeup_small_m0} for
$m_0=56~\GeVcsq$ and $\tb=\sqrt{2}$.  A massless neutralino is excluded by
the chargino, neutralino, and slepton bounds taken together.
\par
The cases $\tb=1$, $\sqrt{2}$, $2$, and $35$ have been analyzed fully, as
illustrated by the examples shown in
Figs.~\ref{closeup_large_m0},~\ref{closeup_medium_m0},
and~\ref{closeup_small_m0}.  The case $\tb=35$ is simpler than the others
because the chargino contours vary less dramatically with $\mu$.  The limit
on $\Mchi$ is obtained by calculating the chargino limit as a function
of~$\mu$ assuming zero efficiency when $\Mcha>\Msnu$ in accordance with
the `blind spot' shown in Fig.~\ref{cha_2_3}. The limit on $\Mcha$ coming
from the~Z line shape is independent of the chargino decay mode, however,
and contributes for $m_0$ in a narrow range around $67~\GeVcsq$.
\par
Fig.~\ref{mchi_m0} displays the lower bound on $\Mchi$ as a function of $m_0$
for fixed values of $\tb$ and $\mu<0$.  The limits for $m_0>70~\GeVcsq$ come
from the charginos and neutralinos, and are nearly independent of $m_0$. Below
some point $m_0\approx 70~\GeVcsq$ the effects of a light sneutrino cause a
precipitous drop in the limit; this drop is more a consequence of effectively
invisible two-body decays than of reduced chargino production.
For $40<m_0<60~\GeVcsq$, the limits follow from the bounds on $M_2$ from
sneutrinos and sleptons, within the $\mu$~range allowed by the charginos and
neutralinos, similar to those depicted in Fig.~\ref{slep_m0_m12}.
In particular, the region of small~$M_2$ and large $m_0$ is important as
it generally overlaps with the bounds from charginos.
The value of $\mu$ giving the lowest $\Mchi$ is used, which generally
differs from that which gives the lowest $M_2$, due to the dependence of
$\Mchi$ on $\mu$.
Sneutrinos dominate for $\tb=2$ and~35, and sleptons dominate for
$\tb=\sqrt{2}$ and~1, hence the slight difference in shape.
\par
For $\tb=1$, 2, and 35 the chargino/neutralino and the slepton/sneutrino
bounds overlap, precluding massless neutralinos.  For~$\tb$ in the
neighbourhood of $\sqrt{2}$, however, a massless $\chi$ is possible,
provided $m_0\approx60~\GeVcsq$, {\em i.e.},
$M_{\tilde{\nu}}\approx 47~\GeVcsq$, $M_{\tilde{\mathrm{e}}_L}\approx
M_{\tilde{\mathrm{e}}_R} \approx 65~\GeVcsq$, and $\mu\approx -60~\GeVcsq$.
The full extent of this massless neutralino window in the $(\mu,m_0)$ plane
is displayed in Fig.~\ref{free_mu}. The larger triangular opening is obtained
allowing $\tb$ to be free, while the dashed contour is obtained fixing
$\tb=\sqrt{2}$.
There is no opening for $\tb<1.2$ or $\tb>1.8$.
This window is caused by the impact of light sneutrinos on the experimental
efficiency for selecting charginos and neutralinos, not by kinematic limits
or a lack of luminosity.
\par
Although the four limits shown in Fig.~\ref{mchi_m0} behave similarly, the
limit on $\Mchi$ for a given $m_0$ plainly varies with $\tb$.  Four values of
$m_0$ have been chosen to illustrate this dependence: $m_0=200~\GeVcsq$,
(Fig.~\ref{mchi_limit}), exemplifies the case in which
sneutrinos and selectrons play a negligible role, while for $m_0=70~\GeVcsq$
their effect is pronounced but not disastrous.  For $m_0=50~\GeVcsq$, the
limit comes entirely from the slepton and sneutrino bounds evaluated for the
$\mu$ ranges allowed by chargino and neutralino bounds.  The value
$m_0=62~\GeVcsq$, giving a relatively weak bound on $\Mchi$, is included
also. The limits on $\Mchi$ for these four fixed values of $m_0$ are shown
in Fig.~\ref{mchi_three_m0_negmu} for $\mu<0$.
\par
Indirect limits on $\Mchi$ from the chargino, neutralino, slepton, and
sneutrino bounds have been derived for $\mu>0$, also, and are shown in
Fig.~\ref{mchi_three_m0_posmu} for the same values of~$m_0$. The limits for
$m_0=70~\GeVcsq$ and $200~\GeVcsq$ are stronger for positive $\mu$
than negative $\mu$ because they come mainly from gauginos.  They also show
little or no dependence on $\tb$.  For $m_0=50~\GeVcsq$, however, a light
sneutrino erodes the chargino limits for large positive $\mu$ and
$\tb\appgt 1.5$.  The strong $\mu$ dependence of $\Mchi$ in this case leads
to a limit weaker than the one obtained for negative~$\mu$.
%
\subsection{Minimal Supergravity Scenario}
\par
The large number of parameters of the MSSM can be reduced by additional
hypotheses, most of which are natural within the framework of
supergravity~\cite{susy}. Here the following further assumptions are made:
The scalar mass parameter $m_0$ is universal, applying for Higgs bosons
and squarks as well as for sleptons and sneutrinos. The trilinear coupling~$A$
is universal for all scalars. Electroweak symmetry breaking proceeds from
radiative corrections induced by the large top quark Yukawa coupling, a
calculation which relates the $\mu$ parameter to the others, up to a sign.
The remaining parameters are therefore $\mhalf$, $m_0$, $A$, $\tb$ and the
sign of $\mu$. The possible sets of values for these parameters are restricted
by the requirements that the top Yukawa coupling should remain perturbative up
to the GUT scale, that none of the scalar particle squared masses should be
negative, and that the LSP should be the lightest neutralino. This set of
assumptions is commonly referred to as ``minimal supergravity.''
\par
In the following, all calculations were performed using the {\sc ISASUSY}
package~\cite{ISASUSY}. The top quark mass was set to $175~\GeVcsq$, except
for values of $\tb$ too small to be consistent with such a mass, in which
case the largest possible top mass was chosen ({\it e.g.} $165~\GeVcsq$ for
$\tb=\sqrt{2}$). The $A$ parameter was set to zero (at the GUT scale).
\par
Exclusion domains in the $(m_0,\mhalf)$ plane are shown in Fig.~\ref{SUGRA}
for $\tb=2$ and~$10$, and for both signs of $\mu$. The dark areas are
forbidden by the requirement that a correct electroweak symmetry breaking be
achieved, or by one of the other theoretical constraints mentioned above.
Substantial domains are excluded by the Z width measurement
($\Msnu>43~\GeVcsq$ and $\Mcha>45~\GeVcsq$).
Large regions are excluded by the Higgs boson searches at \lepa~\cite{Brussels}
({\it e.g.} $m_h>61~\GeVcsq$ for $\tb=2$ and $m_h>45~\GeVcsq$ for $\tb=10$),
particularly for low values of $\tb$ and for negative $\mu$.  These excluded
domains are considerably extended by the chargino searches at \lepb. It turns
out that the constraints from neutralino searches at \lepa~and \lepb\ and the
slepton searches at \lepb\ do not exclude any additional region of parameter
space.
\par
The lower bound on $\Mchi$ as a function of $\tb$ is displayed in
Fig.~\ref{mchi_sugra}.  Large portions of the $(m_0,\mhalf)$ plane are
excluded on theoretical grounds when $\tb<1.2$, resulting in a rapid
rise for the~$\Mchi$ bound as $\tb\goto 1$.  The \lepb\ chargino bound is
effective for~$\tb$ up to~$2.8$, supplemented for negative~$\mu$ by the
Higgs bound, which decreases slowly as~$\tb$ increases.  At $\tb\approx 2.9$,
however, the sneutrino and chargino are almost mass degenerate, and for
positive~$\mu$, the lower limit on $\Mchi$ drops precipitously.  The Higgs
bound prevents such a drop for negative~$\mu$, and for larger $\tb$, the
neutralino bounds come into play, slowly eroding as $\tb\appgt 5$.  The
curve for~$\mu>0$ follows the \lepa\ chargino bound for $\tb>2.9$.
\par
For $\tb=2$, the lowest allowed neutralino mass is $28.5$ and $35~\GeVcsq$
for $\mu<0$ and $\mu>0$, respectively; for $\tb=10$, the limits are~$28$
and~$25~\GeVcsq$.  The absolute lower bound on $\Mchi$ under these
theoretical assumptions is $22~\GeVcsq$.
%
\section{Additional Implications}
\par
The less theoretically constrained analyses of subsections~\ref{a1}
and~\ref{a2} have further implications which are elaborated here.
\par
It has been pointed out recently~\cite{window} that the \lepa~results and the
\lepb\ chargino results do not rule out the so-called ``supersymmetric limit,''
for which $M_1, M_2, \mu \goto 0$ and $\tb\goto 1$.  In this region of the
$(\mu,M_2)$ plane, the~Z does not couple to $\chi$ and $\chip$, and associated
production of $\chip\,\chipp$ and $\chip\,\chippp$ is kinematically suppressed
for $\sqrt{s}\sim 91~\GeV$.  At the higher centre-of-mass energies of \lepb,
however, these channels are open, and the neutralino search of
Ref.~\cite{susy133} and the additional selections detailed in the Appendix
exclude these states, as shown in Fig.~\ref{closeup_101}.  The experimental
upper limit on the cross section near the limit point is 3.0~pb, at 95\% CL,
to be compared to 3.5~pb expected in the MSSM. The range excluded in~$\mu$
survives for small~$m_0$, owing to the dependence of $\Msnu$ on $M_2$.
\par
The possibility that a light gluino might exist has remained experimentally
open, and has attracted theoretical interest~\cite{gl_window}.  Unification of
gauge couplings at the GUT scale implies that the gluino mass is proportional
to $M_2$ (neglecting radiative corrections), so the limits on $M_2$ presented
here imply bounds on the gluino mass.  In particular, within the theoretical
framework assumed in this analysis, a massless gluino is allowed by the \ALEPH\
data only in the region depicted in Fig.~\ref{free_mu}.  However, a new limit
on the selectron mass was published recently by the
{\sc AMY}~Collaboration~\cite{amy}, which extends the region excluded by
\ALEPH~for very light neutralinos, and closes the hole in Fig.~\ref{free_mu}.
\par
Bounds on massive gluinos have been obtained from proton collider
experiments~\cite{d0gluino}. Assuming gauge unification, this limit translates
into $M_2>49~\GeVcsq$. Imposing this constraint in the analysis, the limit
$\Mchi>27~\GeVcsq$ is obtained for large $m_0$, independent of $\tb$.  The
reported gluino limits, however, were obtained for specific values of $\mu$ and
$\tb$, and it is not known how robust they are against variations of these
parameters.
\par
In the slepton sector, the variation of the sneutrino mass with~$\tb$
complements that of the charged sleptons (Eq.~\ref{m_scalers}). Fixing $M_2$,
the sneutrino bound on $m_0$ becomes stronger as $\tb$ increases, while the
slepton bound becomes weaker.  The combined limit turns out to vary little
with $\tb$, making a $\tb$-independent bound in the $(M_2,m_0)$ plane possible:
For each value of $M_2$ and within the ranges of $\mu$ allowed by the
chargino and neutralino searches, the limit in $m_0$ is calculated as a
function of $\tb$.  The lowest value is taken, which occurs for $1<\tb<3$,
giving the result shown in Fig.~\ref{slep_m0_m12_tbind}.
%
\section{Conclusion}
\par
The combination of chargino and neutralino search limits obtained by the
\ALEPH~Collaboration using data acquired at \lepa~and \lepb~excludes a lightest
neutralino with mass less than $12.8~\GeVcsq$, when the sneutrino mass is
greater than~$200~\GeVcsq$. For large values of $\tb$, the lower limit climbs
to $34.1~\GeVcsq$.   The dependence on sneutrino and slepton masses has been
investigated in the context of a SUSY GUT scenario. Although the limits are
weakened when the scalar mass parameter $m_0$ falls below $80~\GeVcsq$, a
massless neutralino is allowed only for restricted ranges of $\tb$, $m_0$, and
$\mu$.  Further theoretical assumptions based on minimal supergravity allow the
absolute lower limit $\Mchi>22~\GeVcsq$. Finally, an excluded region in the
$(M_2,m_0)$ plane independent of $\mu$ and~$\tb$ has been delineated.  All of
these combined limits are far more constraining than those obtained by
considering either \lepa\ or \lepb\ data alone.
%
\section*{Acknowledgments}
\par
Discussions with our theoretical colleagues John~Ellis, Toby~Falk, and
Keith~Olive are gratefully acknowledged.
We thank and congratulate our colleagues from the accelerator divisions for
successfully operating the \lep~machine in this new energy regime so quickly.
We are indebted to the engineers and technicians in all our institutions for
their contributions to the excellent performance of the \ALEPH~detector.
Those of us from non-member states thank {\sc CERN}~for its hospitality.
%
\section*{Appendix}
\par
The analysis of Ref.~\cite{susy133} has been augmented by two sets of
selection criteria which require energetic, isolated photons, in order to
make the results independent of the rate of radiative decay
$\chip\goto\chi\gamma$.  This is especially important in the small~$|\mu|$,
small~$M_2$ regime, where the production of $\chip\chipp$ and $\chip\chippp$
final states is possible at~\lepb.  Definitions of the quantities used and
motivations for the selection criteria are given in Ref.~\cite{susy133}.
\par
The first set of selection criteria is similar to the search for
hadronic events with missing energy, optimized for large mass differences
($\Mcha-\Mchi$).  It is orthogonal to that analysis, however, in that one
photon is required with at least $10~\GeV$ in energy and isolated in so far
as the sum of energies of all energy flow objects inside a cone of $30^\circ$
(and outside a cone of $5^\circ$) around the photon should be less
than $3~\GeV$.
\par
The other requirements were adjusted to give an optimal efficiency for
neutralino events in the small $|\mu|$, small $M_2$ region. The number of
charged particle tracks should be at least seven.  The visible mass of
the event, photon included, should be at least 10\% of $\sqrt{s}$ and no more
than 90\% of $\sqrt{s}$, in order to suppress radiative returns to the~Z.
The total transverse momentum of the event should be at least 10\% of the
visible energy.  The missing momentum vector should point at least
$25.8^\circ$ from the beam axis, and the energy in a $30^\circ$ ``wedge''
around this vector should be no more than 5\% of $\sqrt{s}$.
The acoplanarity angle, calculated from the vector sums of all energy flow
objects in each hemisphere, should be less than $160^\circ$.  Finally, the
direction of a hypothetical neutrino, calculated using energy and momentum
conservation and allowing for initial state radiation along the beam line,
must point at least $25.8^\circ$ from the beam axis.
The expected number of background events is 0.2, coming mainly from
radiative returns to the~Z.
\par
The second set of selectron criteria is based on the two-photon analysis of
Ref.~\cite{multiphotons}.  Some selection criteria were tightened in order to
reduce background expectations, and no attempt was made to incorporate
converted photons.  The event is required to have no charged particle
tracks and a minimum of two photons.
Taking the two most energetic ones, one photon must have energy
greater than $10~\GeV$ and the other energy greater than 5\% of
$\sqrt{s}$.  The angle of the first photon with the beam should be
at least $18.2^\circ$, and the angle of the second, $25.8^\circ$.
The acoplanarity angle of the two photons should be less than $170^\circ$.
Finally, the total energy within $12^\circ$ of the beam axis should be zero.
The background expectation is half an event from
$\mathrm{e}^+\mathrm{e}^-\rightarrow\nu\bar{\nu}\gamma$.
\par
\begin{figure}
\begin{center}
\mbox{\epsfig{file=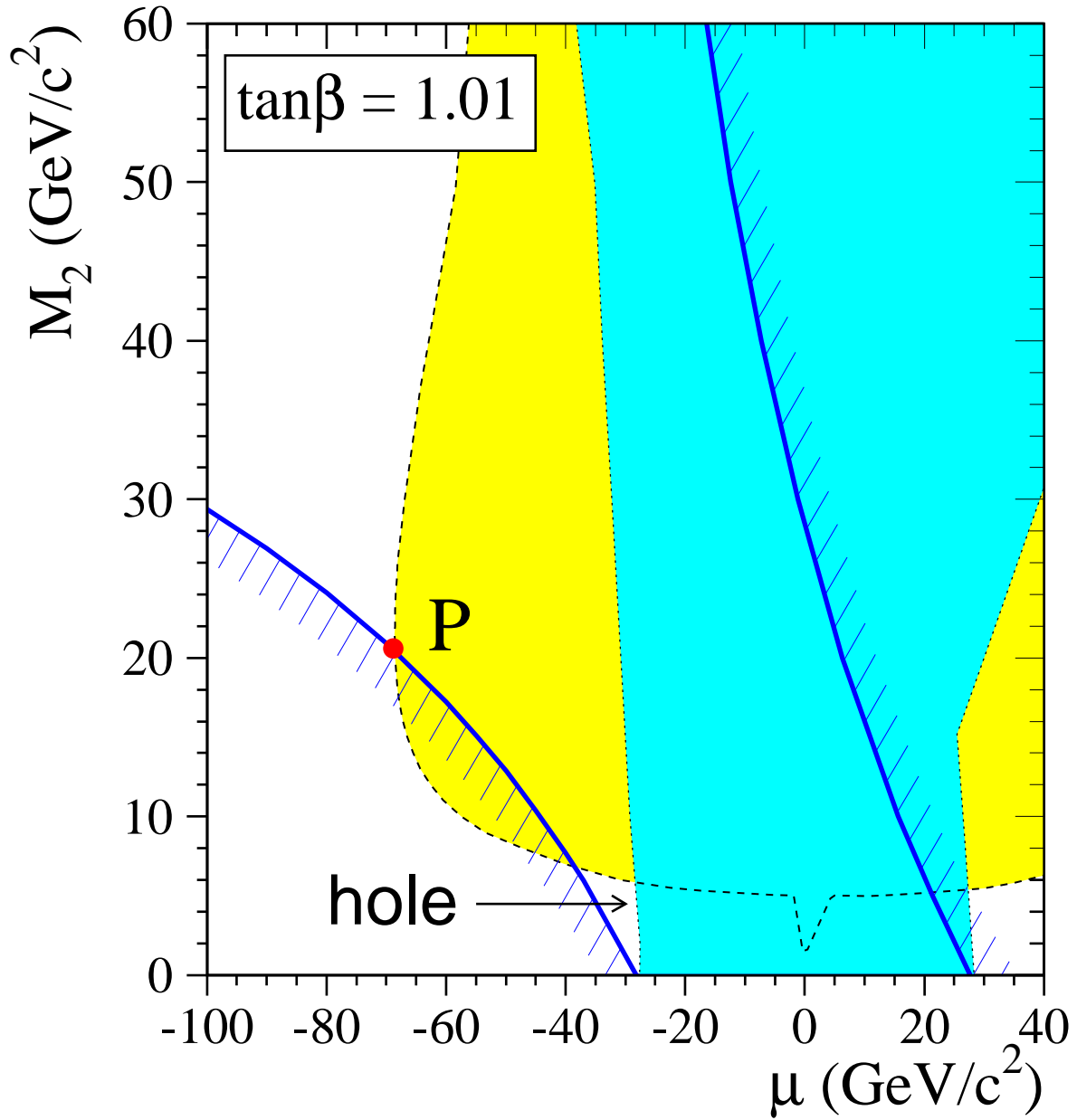,width=16cm%
,bbllx=5mm,bblly=50mm,bburx=165mm,bbury=215mm}}
\end{center}
\caption[.]{\em
Close-up view of the limit contours coming from the
chargino search at \lepb~(heavy solid curve),
from the neutralino search at \lepb~(dark shaded region),
and the direct search for neutralinos at \lepa~(light shaded region),
for $\tan\beta=1.01$ and $\Msnu=200~\GeVcsq$.
A small triangular hole near $\mu=-30~\GeVcsq$ is evident.
\label{closeup_101} }
\end{figure}
\begin{figure}
\begin{center}
\mbox{\epsfig{file=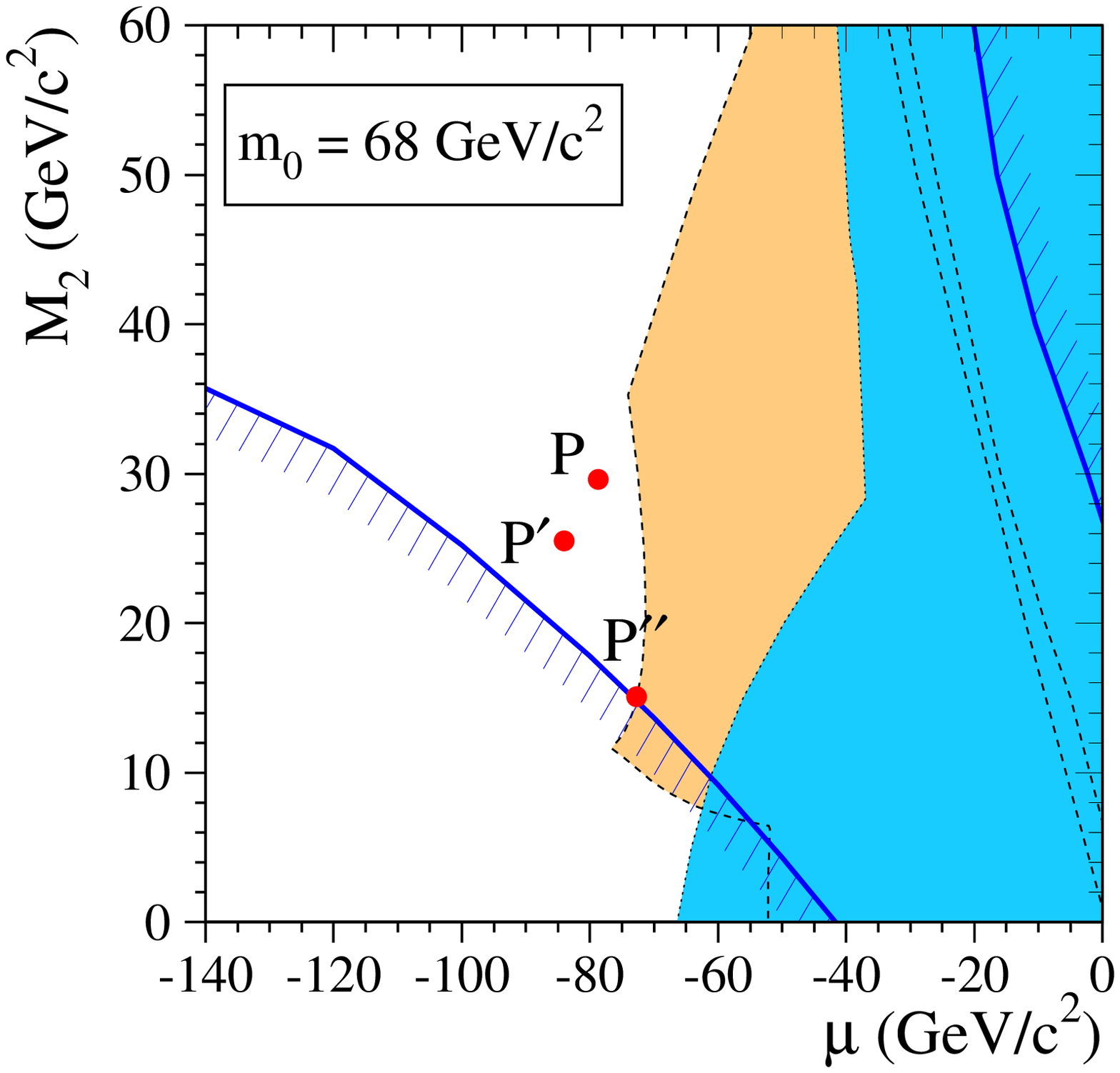,height=16.5cm%
,bbllx=20mm,bblly=50mm,bburx=155mm,bbury=225mm}}
\end{center}
\caption[.]{\em
Close-up view of limit contours for $\tb=\sqrt{2}$ and $m_0=68~\GeVcsq$.
Contours have the same meaning as in Fig.~\ref{closeup_large_m0}.
See the text for an explanation of the points~{\sc P},
{\sc P}$^{\prime}$, and~{\sc P}$^{\prime\prime}$.
\label{closeup_medium_m0} }
\end{figure}
\begin{figure}
\begin{center}
\mbox{\epsfig{file=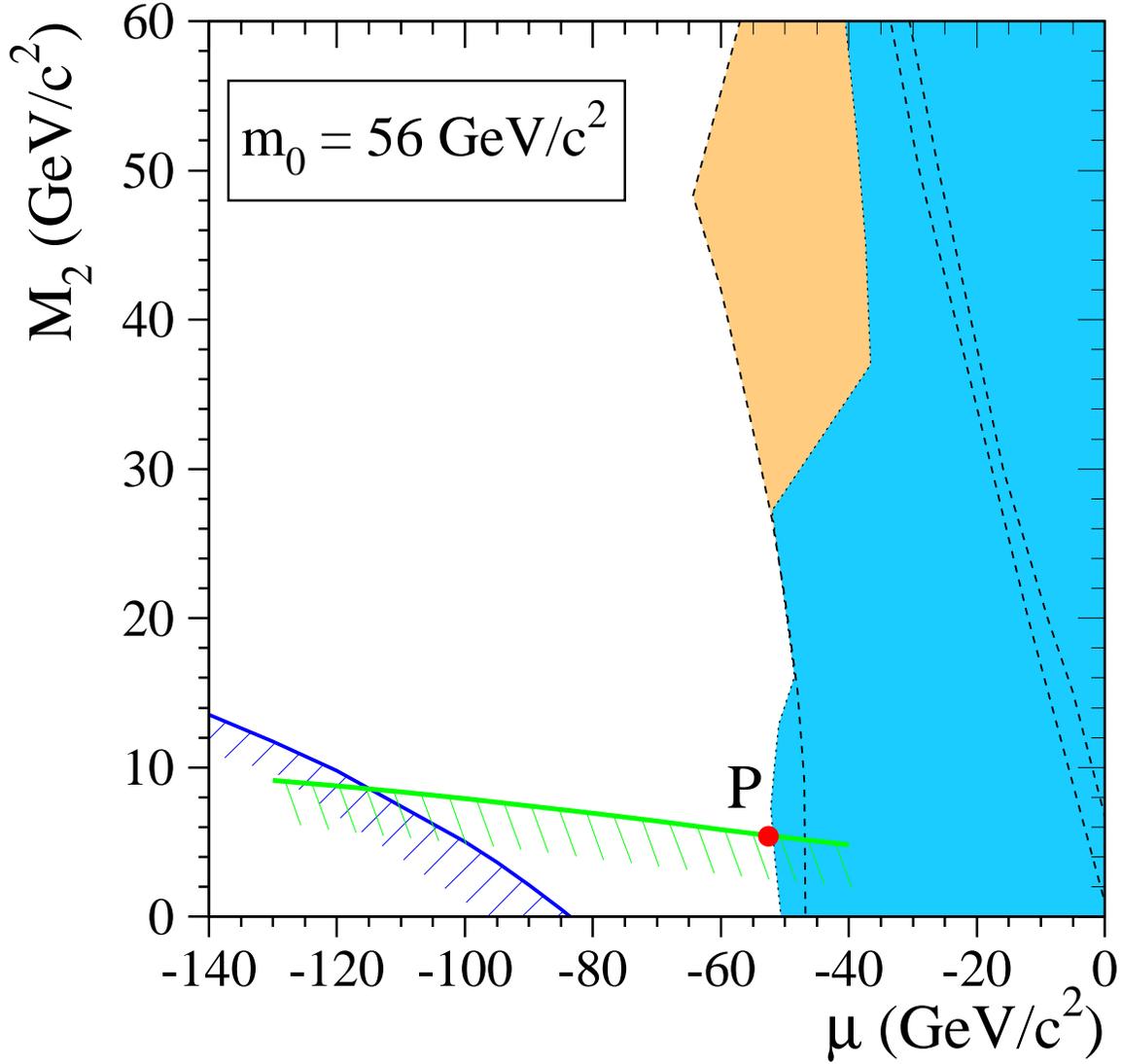,height=16.5cm%
,bbllx=20mm,bblly=50mm,bburx=155mm,bbury=225mm}}
\end{center}
\caption[.]{\em
Close-up view of limit contours for $\tb=\sqrt{2}$, for $m_0=56~\GeVcsq$.
They have the same meaning as in Fig.~\ref{closeup_large_m0}.  The
nearly horizontal solid line passing through point~{\sc P} is the bound
derived from the slepton search.
\label{closeup_small_m0} }
\end{figure}
\begin{figure}
\begin{center}
\mbox{\epsfig{file=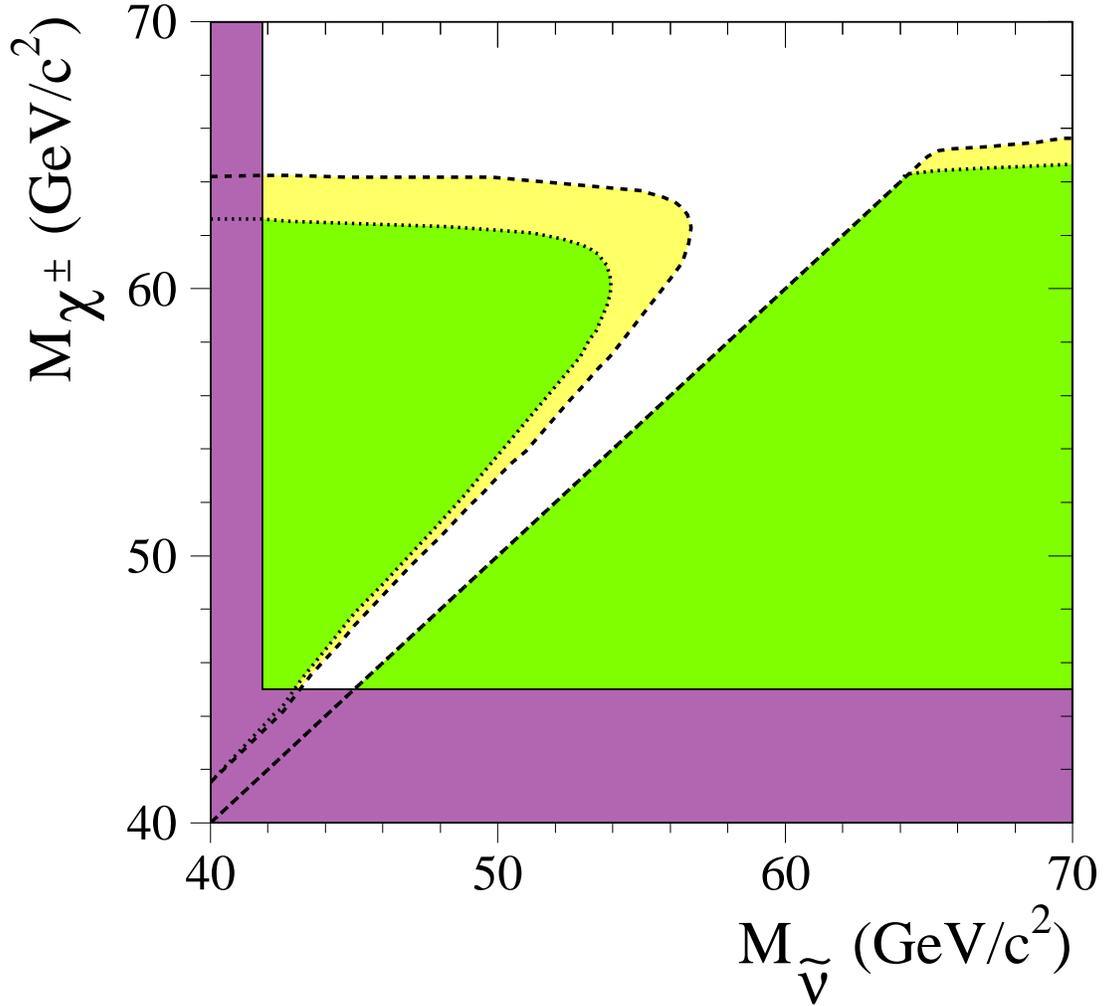,width=16cm%
,bbllx=-10mm,bblly=0mm,bburx=170mm,bbury=160mm}}
\end{center}
\caption[.]{\em
Exclusion contour for charginos as a function of $M_{\snu}$.
The case of the near gaugino region ($\mu=-100~\GeVcsq$) is represented
by the medium dark region, which is extended to the light region for the far
gaugino region ($\mu=-500~\GeVcsq$).
The limits from \lepa~are indicated by the darkest region.
\label{cha_2_3} }
\end{figure}
\begin{figure}
\begin{center}
\mbox{\epsfig{file=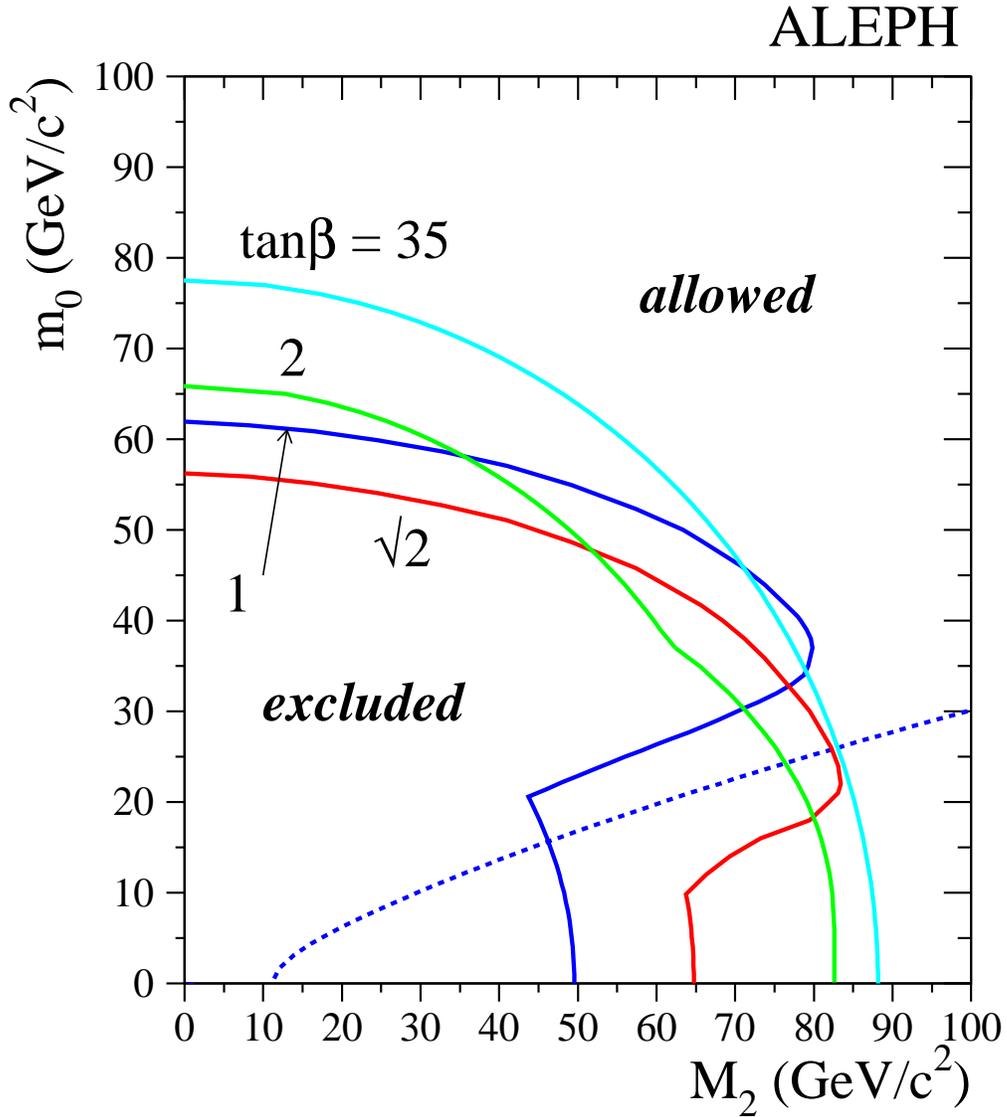,height=16.5cm%
,bbllx=12mm,bblly=30mm,bburx=160mm,bbury=235mm}}
\end{center}
\caption[.]{\em
The solid curves show the region in the $(M_2,m_0)$ plane excluded by the
combined slepton and sneutrino limits for fixed values of $\tb=1$, 
$\sqrt{2}$, $2$, and~$35$, computed for $\mu<0$.  The dashed curve shows the
theoretical limit $M_{\tilde{\ell}_R} > \Mchi$ for $\tb=1$.
\label{slep_m0_m12} }
\end{figure}
\begin{figure}
\begin{center}
\mbox{\epsfig{file=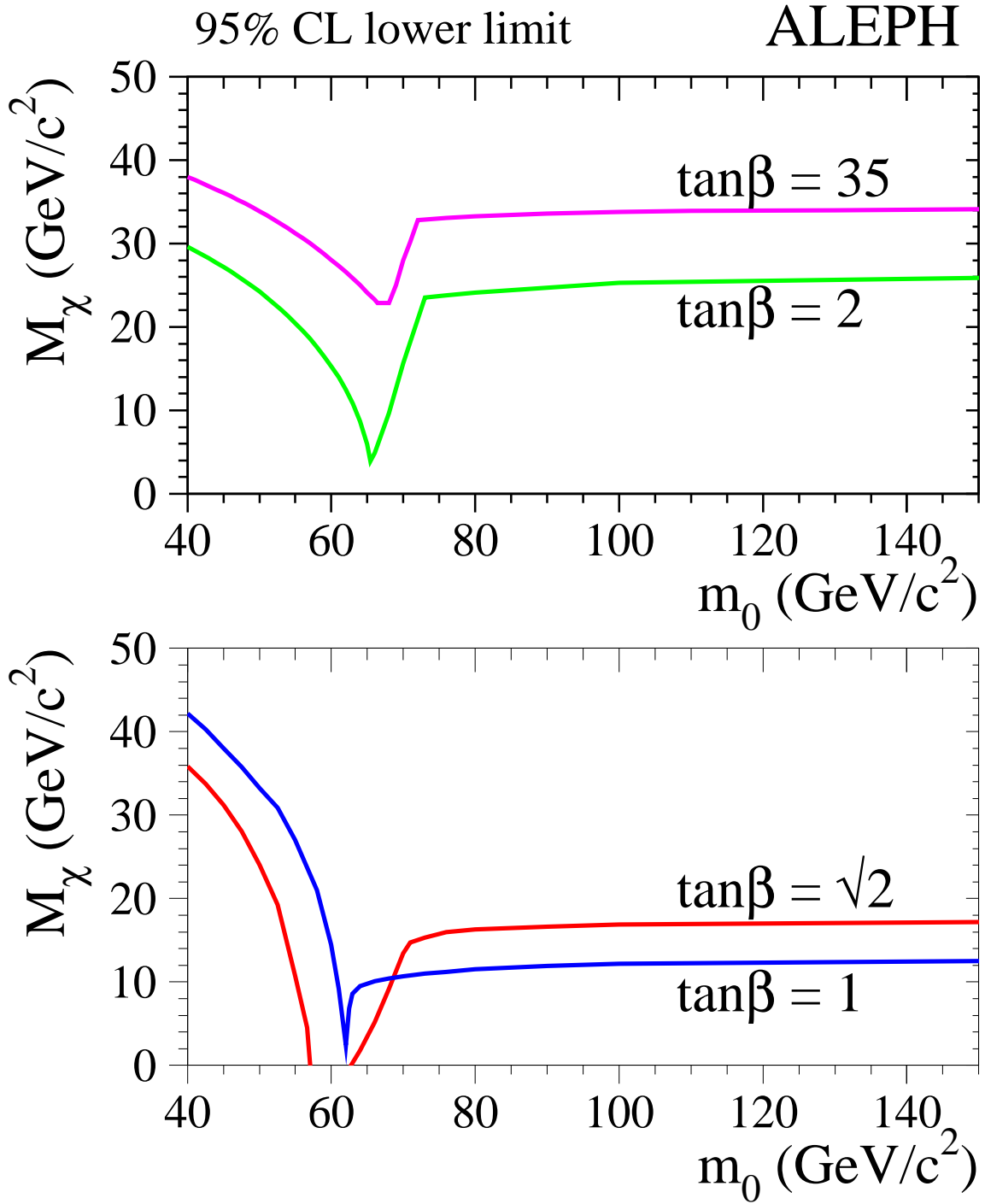,height=18cm%
,bbllx=-10mm,bblly=30mm,bburx=160mm,bbury=217mm}}
\end{center}
\caption[.]{\em
Limit on $\Mchi$ as a function of $m_0$ for $\tb=35$, $2$, $\sqrt{2}$,
and~$1$, restricting $\mu<0$.  For $m_0\appgt65~\GeVcsq$, the limit comes
from the chargino and neutralino searches, while for $m_0\applt65~\GeVcsq$,
it comes from the slepton and sneutrino constraints.
\label{mchi_m0} }
\end{figure}
\begin{figure}
\begin{center}
\mbox{\epsfig{file=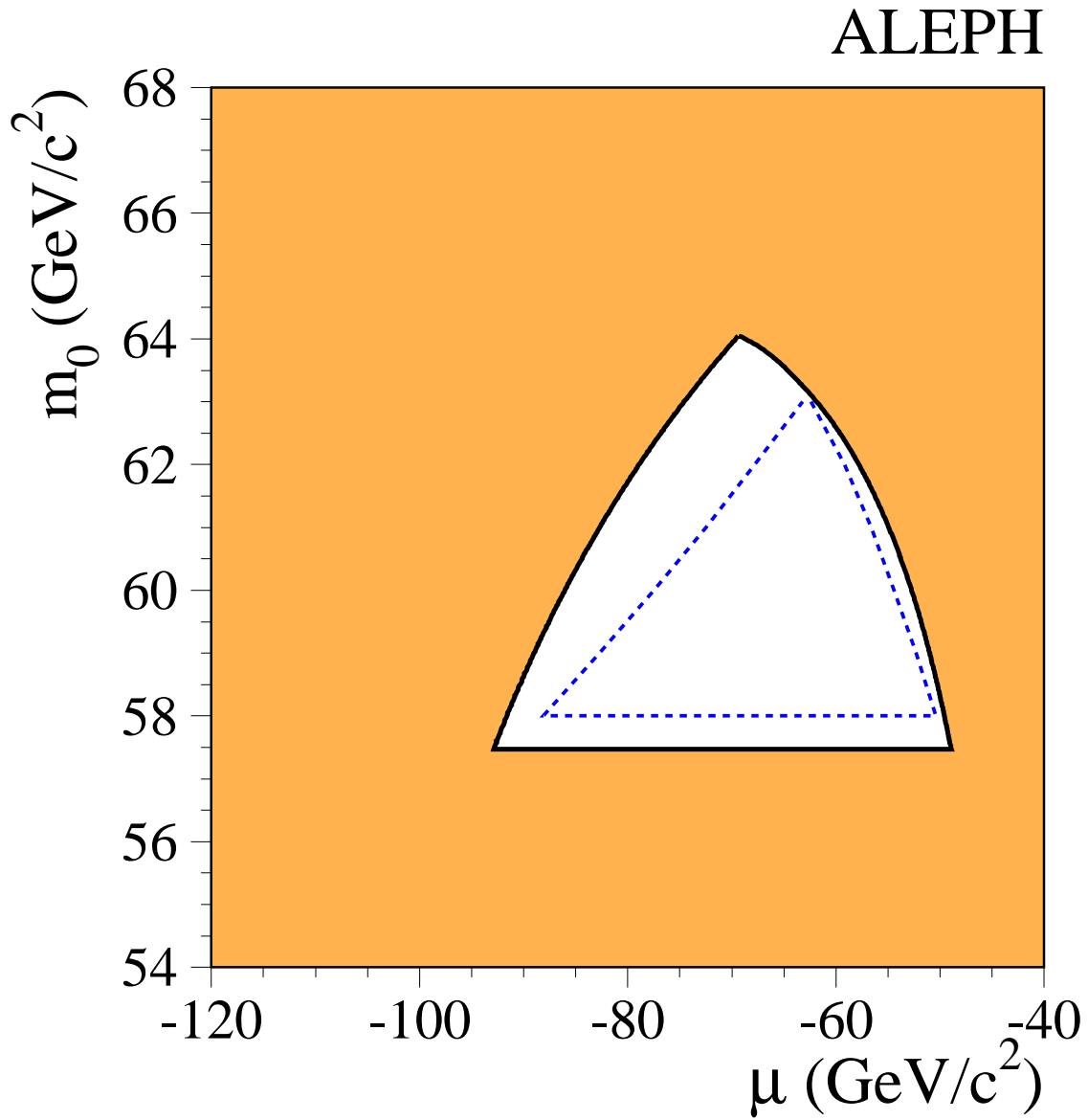,height=16cm%
,bbllx=1mm,bblly=55mm,bburx=165mm,bbury=225mm}}
\end{center}
\caption[.]{\em
Opening in the $(\mu,m_0)$ plane in which $M_2=0$, hence a massless neutralino,
is allowed.  The larger triangular window is obtained allowing $\tb$ to be
free; the dashed curve shows the window for $\tb=\sqrt{2}$.
\label{free_mu} }
\end{figure}
\begin{figure}
\begin{center}
\mbox{\epsfig{file=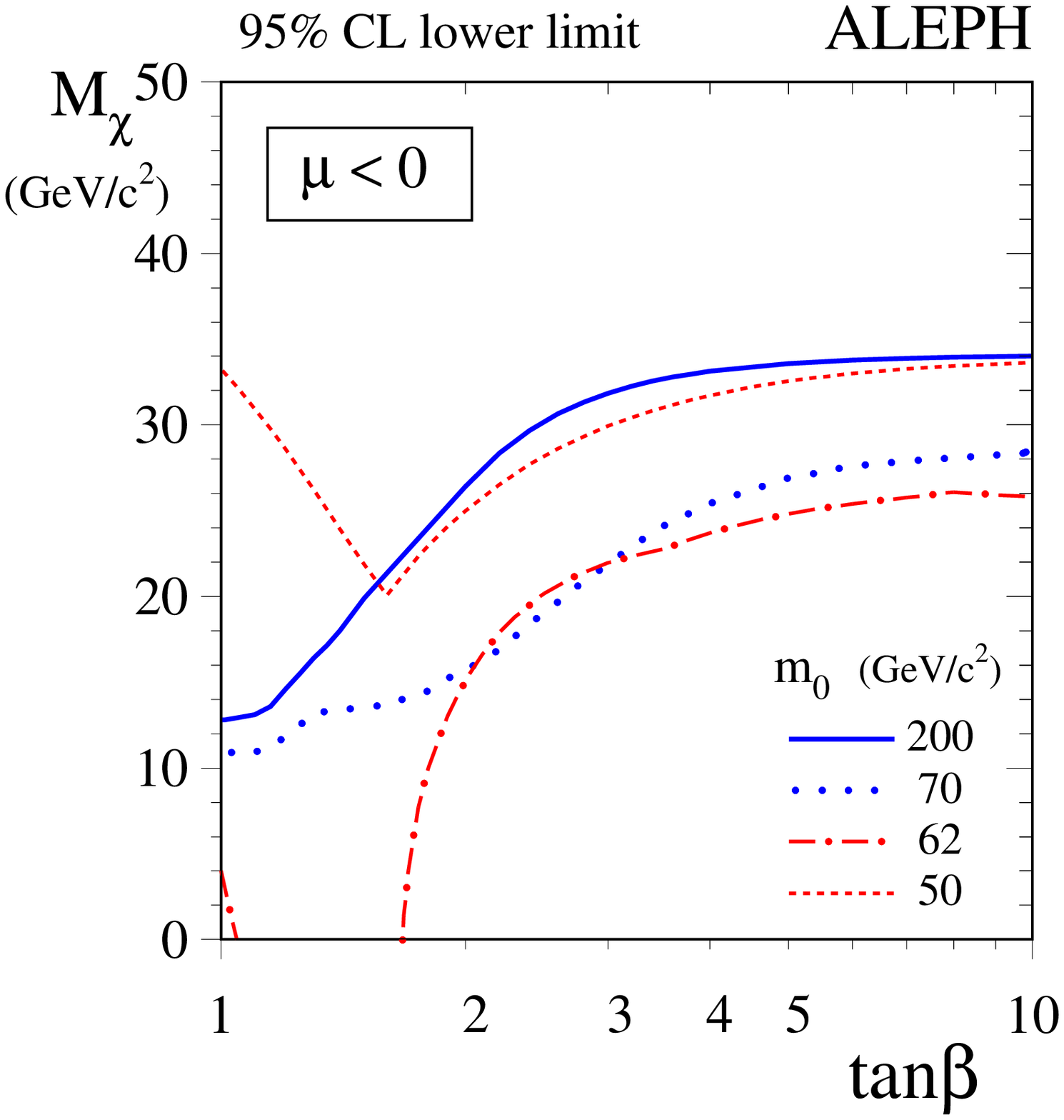,height=16cm%
,bbllx=1mm,bblly=55mm,bburx=165mm,bbury=225mm}}
\end{center}
\caption[.]{\em
For $\mu<0$, the lower limit on the mass of the lightest
neutralino as a function of $\tb$, for $m_0=200$, $70$, $62$,
and $50~\GeVcsq$.
\label{mchi_three_m0_negmu} }
\end{figure}
\begin{figure}
\begin{center}
\mbox{\epsfig{file=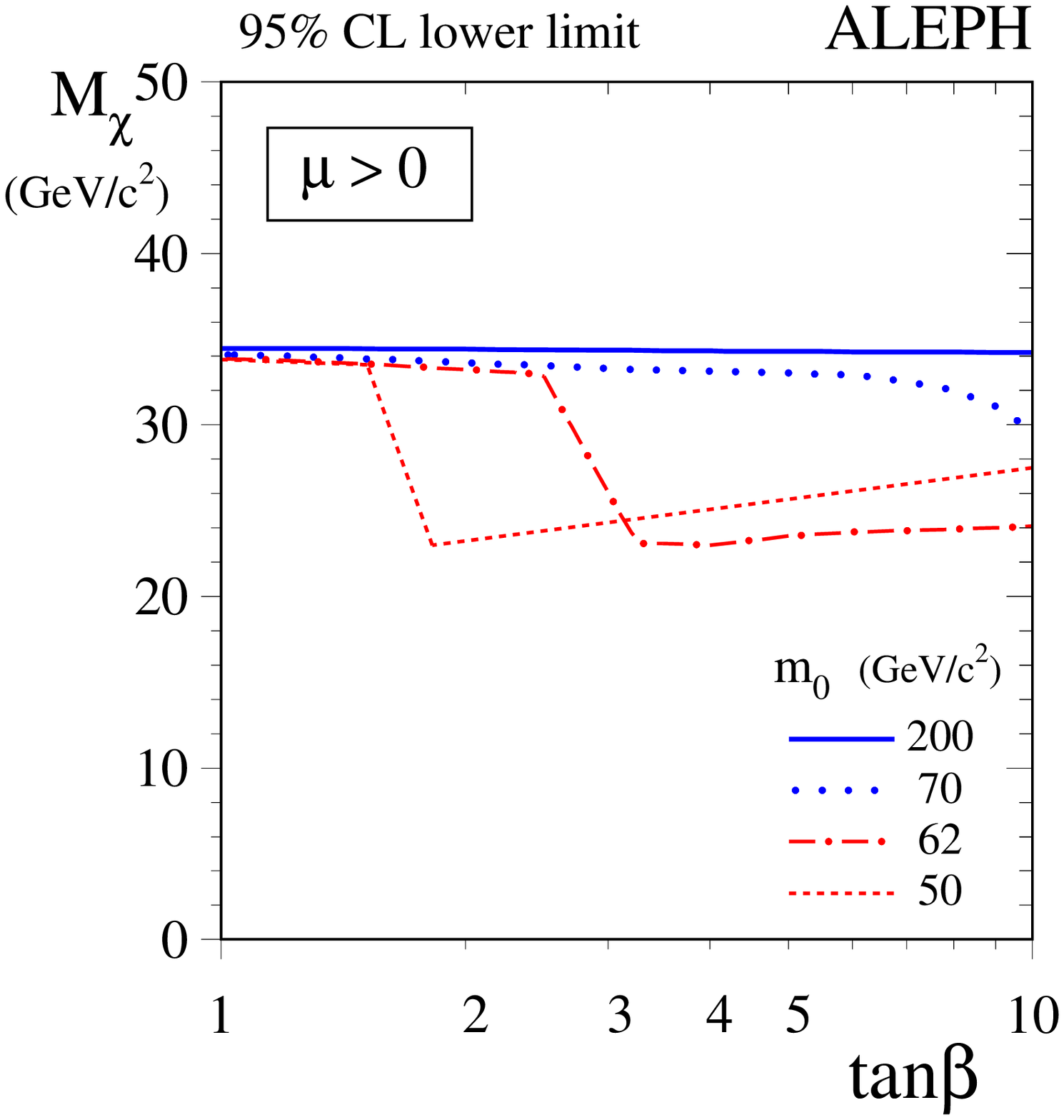,height=16cm%
,bbllx=1mm,bblly=55mm,bburx=165mm,bbury=225mm}}
\end{center}
\caption[.]{\em
For $\mu>0$, the lower limit on the mass of the lightest
neutralino as a function of $\tb$, for $m_0=200$, $70$, $62$,
and $50~\GeVcsq$.
\label{mchi_three_m0_posmu} }
\end{figure}
\begin{figure}
\begin{center}
\mbox{\epsfig{file=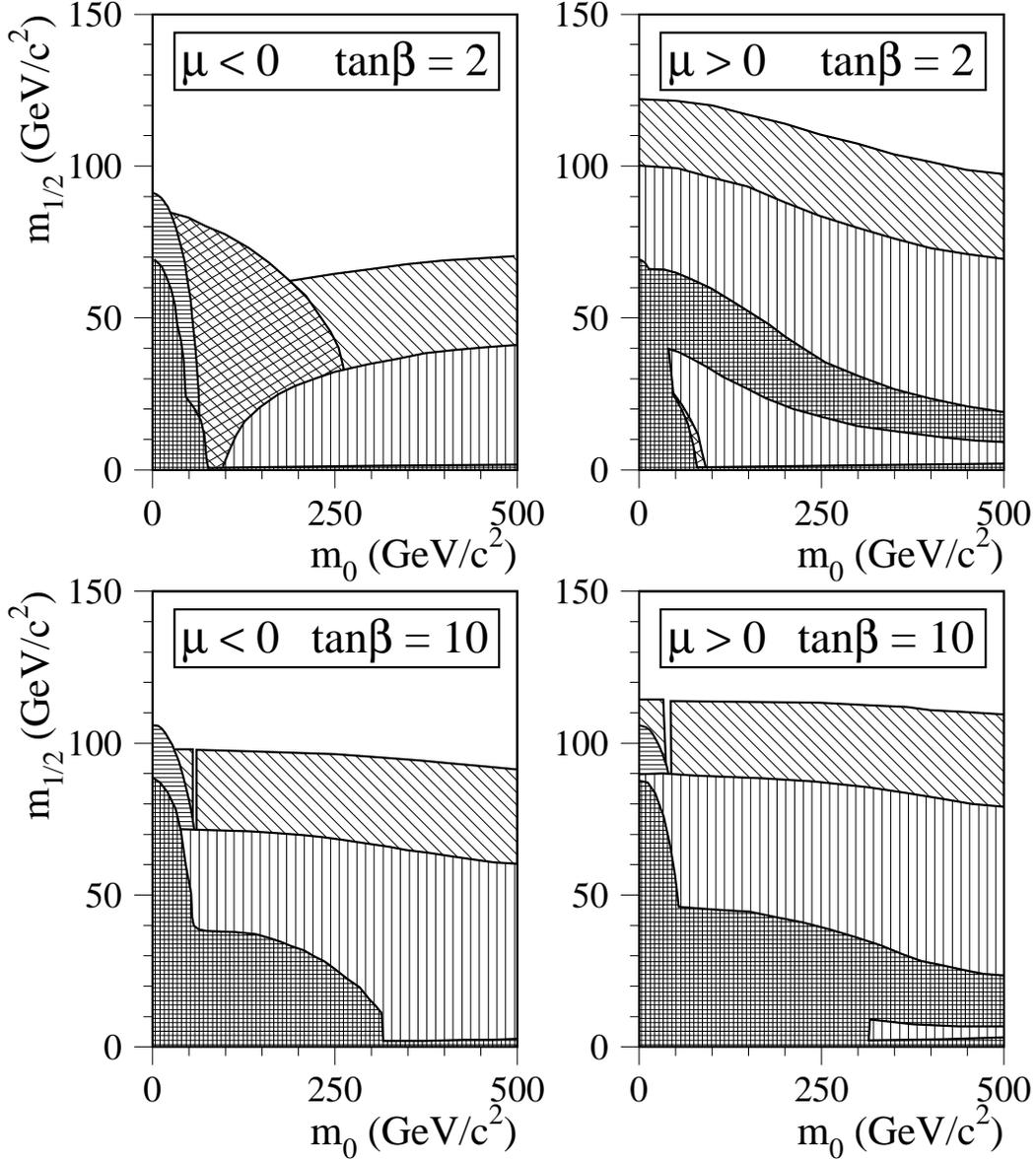,height=16cm%
,bbllx=1mm,bblly=35mm,bburx=175mm,bbury=230mm}}
\end{center}
\caption[.]{\em
Excluded domains in the $(m_0,\mhalf)$ plane for $\tb=2$~(top) 
and~$10$ (bottom), and for $\mu<0$ (left) and $\mu>0$ (right). The
dark shaded regions are theoretically excluded. The vertical, horizontal,
crossed and slanted hatched regions are excluded by charginos at \lepa,
by sneutrinos at \lepa, by Higgs bosons at \lepa~and by charginos
at \lepb, respectively.
\label{SUGRA}}
\end{figure}
\begin{figure}
\begin{center}
\mbox{\epsfig{file=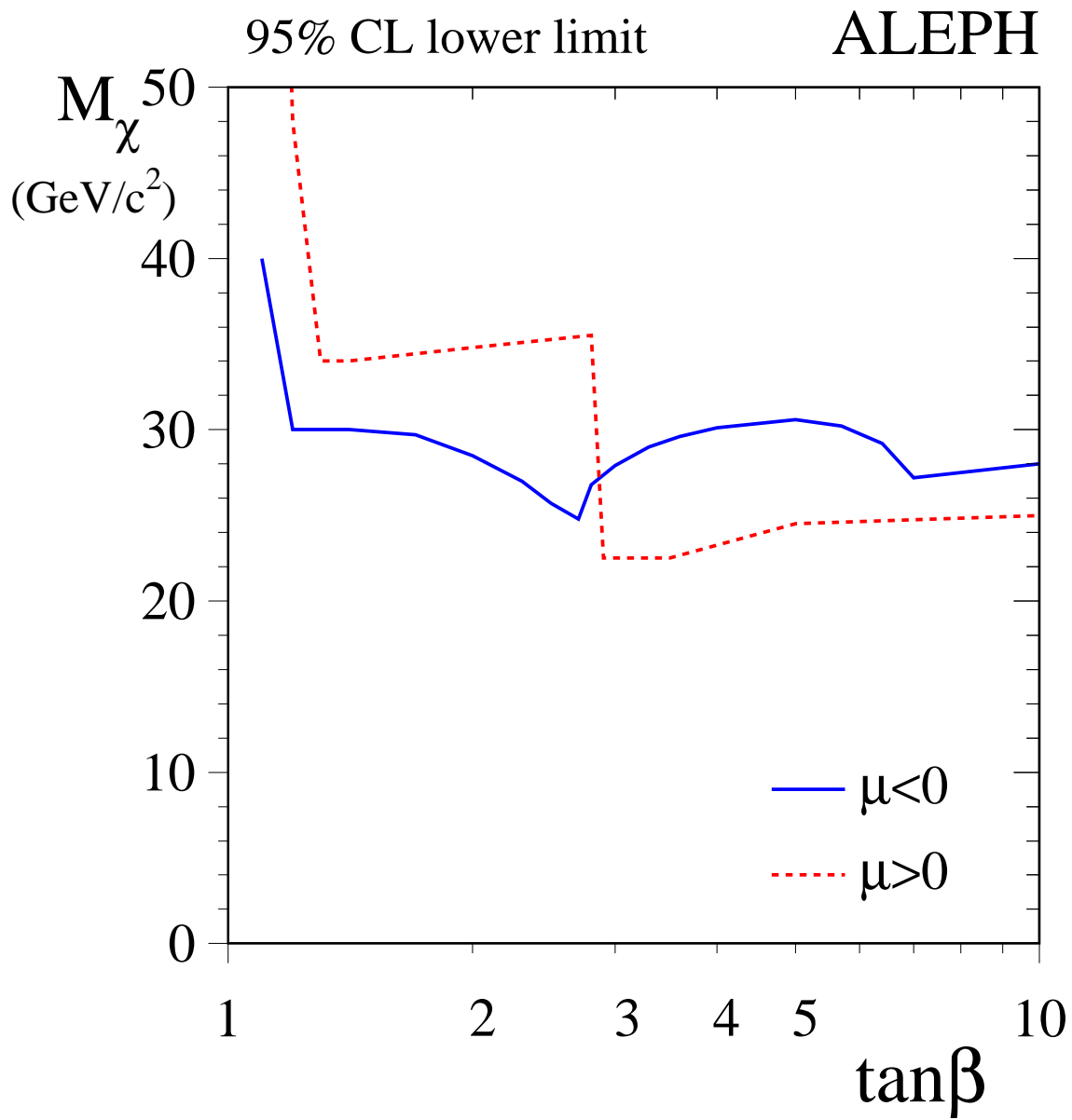,height=16cm%
,bbllx=1mm,bblly=55mm,bburx=165mm,bbury=225mm}}
\end{center}
\caption[.]{\em
Lower limit on the mass of the lightest neutralino as a function
of $\tb$, obtained in the context of minimal supergravity.
\label{mchi_sugra} }
\end{figure}
\begin{figure}
\begin{center}
\mbox{\epsfig{file=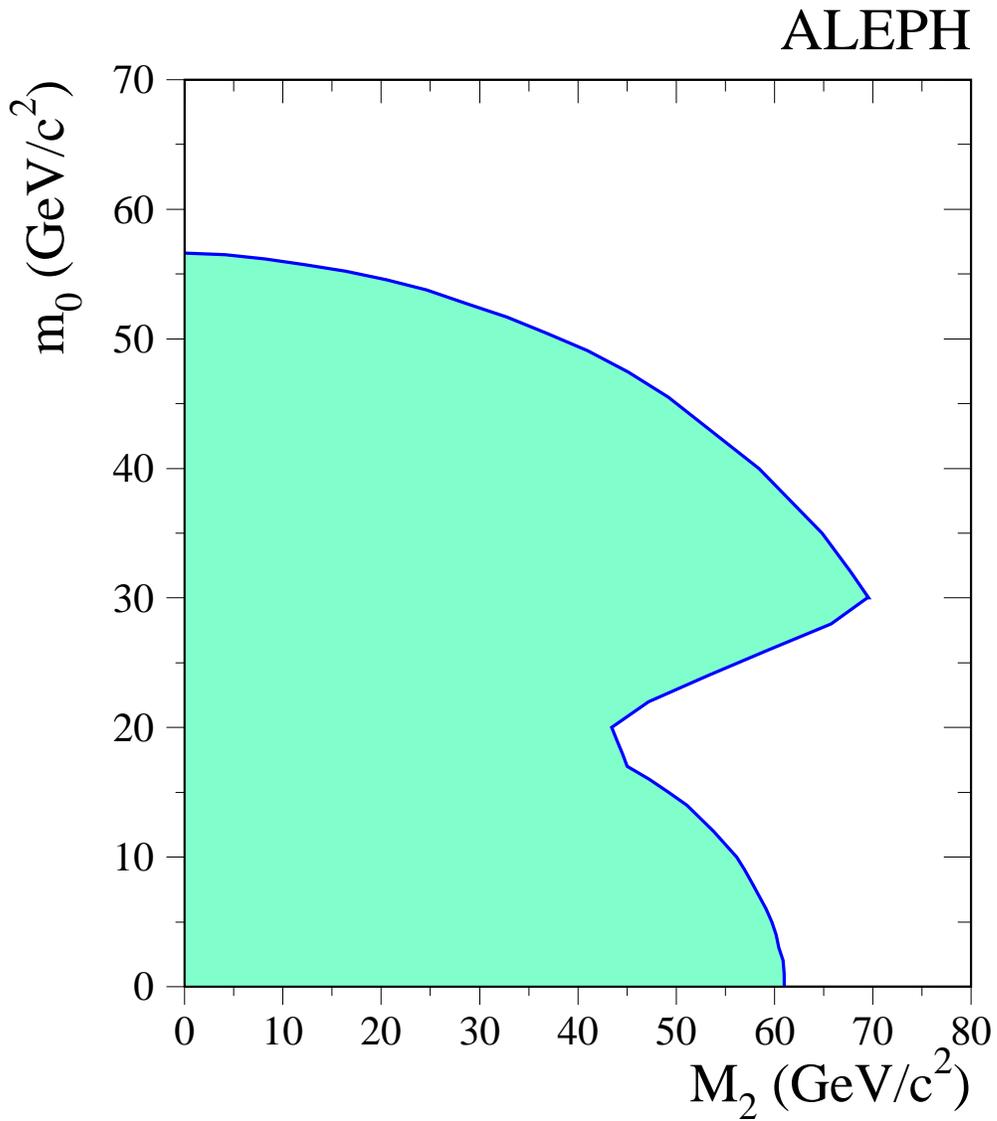,height=16.5cm%
,bbllx=12mm,bblly=30mm,bburx=160mm,bbury=235mm}}
\end{center}
\caption[.]{\em
Region in the $(M_2,m_0)$ plane excluded by charginos, neutralinos,
sneutrinos, and sleptons, valid for all $\mu$ and $\tb$.
\label{slep_m0_m12_tbind} }
\end{figure}

The efficacy of these selections can be judged from two test cases.
The first one ($\mu=-1~\GeVcsq$, $M_2=1~\GeVcsq$, $\tb=1.01$)
is near the supersymmetric limit, and the first two neutralinos are nearly
massless.  The analysis of Ref.~\cite{susy133} accepts a signal cross section
of 0.32~pb, to which the two selections described above add 0.23~and 0.11~pb,
respectively; the total number of events expected would be 3.5.
The second point in parameter space ($\mu=-25~\GeVcsq$, $M_2=1~\GeVcsq$,
$\tb=1.01$)  is close to the boundary in Fig.~\ref{closeup_101}.
The new selections add~0.32 and 0.18~pb to the 0.10~pb accepted by the old
analysis, giving an expected signal of 3.1~events.
These yields are estimated taking $m_0$ to be large; they would increase
for smaller $m_0$.

%
%
%
%
%
%
%
%
%
%
%
%
%
%
%
%
\end{document}